\documentclass[fleqn,usenatbib]{mnras}

\usepackage{newtxtext,newtxmath}
\usepackage{color,soul}
\usepackage{hyperref}

\usepackage[T1]{fontenc}

\DeclareRobustCommand{\VAN}[3]{#2}
\let\VANthebibliography\thebibliography
\def\thebibliography{\DeclareRobustCommand{\VAN}[3]{##3}\VANthebibliography}

\usepackage{graphicx}	
\usepackage{amsmath}	
\usepackage{fix-cm}     

\renewcommand{\hl}[1]{#1}
\makeatletter
\let\textbf@original\textbf
\renewcommand{\textbf}[1]{#1}
\newcommand{\textbftwo}[1]{\textbf@original{#1}}
\renewcommand{\textbftwo}[1]{#1}
\makeatother

\title[Formation of Accretion Disks]{Formation of Accretion Disks around Intermediate-Mass Black Holes in Ultra Compact Dwarf Galaxies and Nuclear Stellar Clusters}

\author[S. Darr and  K. Bekki]{
Sebastian Darr,$^{1}$\thanks{E-mail: 23399856@uwa.student.edu.au}
Kenji Bekki$^{1}$
\\
$^{1}$ICRAR M468, The University of Western Australia, 35 Stirling Hwy, Crawley Western Australia 6009, Australia
}

\date{Accepted XXX. Received YYY; in original form ZZZ}

\pubyear{\the\year{}}

\begin{document}
\label{firstpage}
\pagerange{\pageref{firstpage}--\pageref{lastpage}}
\maketitle

\begin{abstract}
Recent observations have found many ultra-compact dwarf galaxies (UCDs) that contain massive black holes (MBHs) typically consisting of $10\%-15\%$ of their total mass, with some as high as $40\%$. However, whether UCDs initially started with MBHs or grew from intermediate-mass black holes (IMBHs) is unclear. We propose a mechanism in which AGB stars have their gas ejected which forms an accretion disk around an IMBH to fuel BH growth. We used smooth particle hydrodynamical simulations to (i) model the creation and evolution of gas disks that can finally evolve into accretion disks and to investigate (ii) how the properties of the gas disk depend on the model. We find that the amount of mass in the disk increases more with UCD mass than with IMBH mass. For lower-mass IMBHs of $100 \text{ M}_\odot$, a majority of the gas mass present can be trapped in a disk, leading to the idea that lower-mass BHs can utilise the proposed mechanism. The main mechanism working against accretion disk formation is the creation of new stars within the disk, not just through the removal of gas but through gravitational disruption as well. Finally, an estimation of the total BH growth provides a list of observed UCDs which could have formed MBHs using the proposed mechanism which can be useful for verifying the proposed mechanism. Future simulations aimed at modelling the formation of such UCDs can be employed to provide a full picture of the results presented in this paper.
\end{abstract}




\section{Introduction}
Ultra-Compact Dwarf galaxies (UCDs) are distinct systems characterised by their high stellar densities \(\approx 10^7 M_\odot\) and half-light radii \(\gtrsim 10pc\) \citep{brodie2011relationships, mayes2021testing}. UCDs have a typical age of \(\gtrsim 8 \ Gyrs\) \citep{zhang2018stellar} and \([Z/H]\) \(\approx -0.83\) dex \citep{paudel2023creation}. These properties set UCDs apart, establishing them as a unique class of stellar systems that occupy an intermediary position between regular dwarfs and globular clusters (GCs). These exotic stellar objects were discovered by \cite{hilker1999central} and \cite{Drinkwater_Jones_Gregg_Phillipps_2000} through spectroscopic surveys of the Fornax cluster. The objects they found were compact with luminosities similar to those of regular dwarfs. The stellar population of UCDs are primarily old with little to no star formation (SF).
 
Since the discovery of UCDs, two main formation mechanisms have been proposed. Firstly, UCDs are \hl{\textbf{proposed to be}} the remnant core of galaxies that undergo tidal disruptions \citep{Bekki_2001, bekki03, gregg2003galaxy}. This mechanism employs the idea that gas and stars outside the core are stripped more easily. After tidal disruption events occur, usually via cluster environments, only the dense core is left behind. From hydrodynamical simulations of tidally stripped galaxies, this formation path results in more massive UCDs (\(> 10^7 M_\odot\)) \citep{10.1093/mnras/staa3731}.

The second formation mechanism \hl{\textbf{proposes that}} UCDs are large examples of GCs. To form sufficiently large GCs, they require a history of mergers \citep{FellhauerKroupa2002, Mieske2012} \hl{\textbf{and}} must occupy the brighter end of the GC luminosity function \citep{mieske2002ultra}. From these formation mechanisms, interactions with other objects are important in UCD formation, \hl{\textbf{leading to the prediction that they must form in high interaction environments such as in clusters.}} However, recent studies have broadened the search range and found promising UCD candidates outside cluster environments \citep{Saifollahi_2021, Rhode_2023}, suggesting that UCDs were found primarily in cluster environments due to sampling bias from high-density regions. Further study of UCDs in low-density environments can change our understanding of the history of UCD formation.

The dynamical mass-to-light ratios (M/L) of UCDs are \(\approx 50\%\) higher than predicted compared to other stellar models \citep{Dabringhausen2008}. It has been determined that for large UCDs (\(M > 10^7 M_{\odot}\)), the increase in M/L can be attributed to a central black hole (BH) that consists of \(10\%\) to \(15\%\) of the total mass of the UCD \citep{mieske2013central}. Hence, massive black holes (MBHs) \hl{\textbf{are predicted to}} exist around \(10^6 \ M_{\odot}\) in the core of UCDs. \hl{\textbf{MBHs are typically defined in the mass range of $>10^5 \text{ M}_{\odot}$} and occupies the lower end of the supermassive BH mass range.} Many previous studies have confirmed MBHs in UCDs that match these predictions, such as a \(2.1 \times 10^7 M_{\odot}\) MBH within M60-UCD1 \citep{seth2014supermassive}. In addition, a \(4.4 \times 10^6 M_{\odot}\) MBH in VUCD3 and a \(5.8 \times 10^6 M_{\odot}\) MBH in M59c0 \citep{ahn2017detection} have been confirmed. These MBHs contribute \(15\%\), \(13\%\) and \(18\%\) of their parent systems' total mass, respectively. For MBHs that contain a large fraction of the total UCD mass, their influence on the UCD's evolution is paramount. Understanding the evolution of the central MBH can \hl{\textbf{provide insight into the many unanswered questions of UCDs, such as their formation histories and why they have certain properties.}}

Intermediate-Mass black holes (IMBHs) are believed to be the link between stellar-mass BHs and supermassive BHs and are defined in the mass range of \(100 \ M_{\odot}\) to \(10^5 \ M_{\odot}\). Detection of IMBHs has been a challenge until recently, with detections steadily increasing through radio and X-ray interferometry. Recent findings have constrained the mass of IMBH candidates to \(10^{4 - 5} \ M_{\odot}\) \citep{chilingarian2018population, yang2023intermediate, Peißker_2024}. Some spectroscopy studies have found IMBH candidates in GCs \citep{gerssen2002hubble} with masses of \(\approx10^3 \ M_{\odot}\). More recently, an IMBH candidate has been found in the centre of \(\omega\) Centauri with a lower mass bound of \(8200 M_{\odot}\) \citep{haberle2024fast}. Additionally, X-ray bursts have been found within UCDs corresponding to AGN activity from IMBHs rather than neutron stars. Such X-ray bursts have been measured from a UCD companion of NGC 5128 \citep{irwin2016ultraluminous}. 

It is unclear how and why MBHs of such high mass fractions can exist in these stellar systems. This question stems from the lack of understanding of the formation history of UCDs and unrevealed aspects of galaxy evolution. The existence of UCDs with MBHs that contain $\approx10\% -15\%$ of their total mass implies that the evolutionary history of MBHs has had a large effect on the history of the UCD. Therefore, studying the history and \hl{\textbf{potential growth}} of IMBHs in UCDs will reveal unknown aspects of IMBH growth and UCD formation. In this paper, we study IMBH growth in UCDs though a proposed mechanism via \hl{\textbf{smooth-particle hydrodynamical (SPH)}} simulations. We utilised the accepted consideration that BHs grow during AGN activity, which occurs when a BH forms an accretion disk \hl{\textbf{and mass is able to more easily fall into the BH}}. We propose that AGB stars have their gas stripped \hl{\textbf{via stellar winds and accreted by a central IMBH, facilitating BH growth by bringing the AGB gas near the BH}}. 

Previous work has similarly studied the growth of BHs through stellar winds of AGB stars. Specifically, \cite{norman1988evolution} investigated the growth of BHs from stellar winds, some originating from AGB stars, using analytical models. Others have studied accretion disks formed by binaries \citep{chen2017mass, chen20203d}, the ideas of which can be used to model BH growth. However, the formation of accretion disks around IMBHs in UCDs has not been \hl{\textbf{explicitly}} studied in previous simulations or analytical models. Our simulations thus aim to provide confirmation for accretion disk formation around IMBHs in UCDs via \hl{\textbf{stellar winds}} and provide information on the environment and circumstances required for accretion disk formation.

The purpose of this paper is to investigate the conditions necessary to form massive accretion disks around an IMBH by conducting \hl{\textbf{SPH}} simulations of AGB gas particles in the presence of an IMBH's \hl{\textbf{potential}}. The simulations will start with an existing IMBH between the mass range of \(3\times10^3 \ M_{\odot}\) to \(10^5 \ M_{\odot}\) and begin with the \hl{\textbf{AGB stars having their gas ejected and then model the dynamics and properties of AGB gas particles over time}}. Before simulations are performed, an estimate of the total AGB ejecta must be calculated using the initial mass function (IMF). Here, we consider a top-heavy slope $\alpha=1.4$ \citep{kroupa2002imf} and a standard slope $\alpha=2.35$ \citep{salpeter1955luminosity}. \hl{\textbf{The simulations will utilise the IMF to provide accretion disks that form with a reasonable amount of available AGB gas. The results of such simulations can be compared with observational UCDs to potentially explain their MBHs in terms of parameters and environments necessary for their existence. Other simulations which set the available gas at chosen values are employed to study the roll of individual AGB stars on the formation of accretion disks.}}

In Section \ref{DeterminingInitialParameters}, we will discuss the method used to calculate \hl{\textbf{the}} number of AGB stars for a given UCD's stellar mass. We also provide a method for \hl{\textbf{estimating the smallest initial mass a BH can be to produce present day observed MBHs in UCDs.}} An overview of the \hl{\textbf{SPH}} simulations used is also provided. The results of the study are collected in Section \ref{Discussion} and a discussion of the resulting analysis will be found in Section \ref{finaldiscussion}.

\section{Model}\label{DeterminingInitialParameters}
\begin{figure*}
    \centering
    \includegraphics[width=\textwidth]{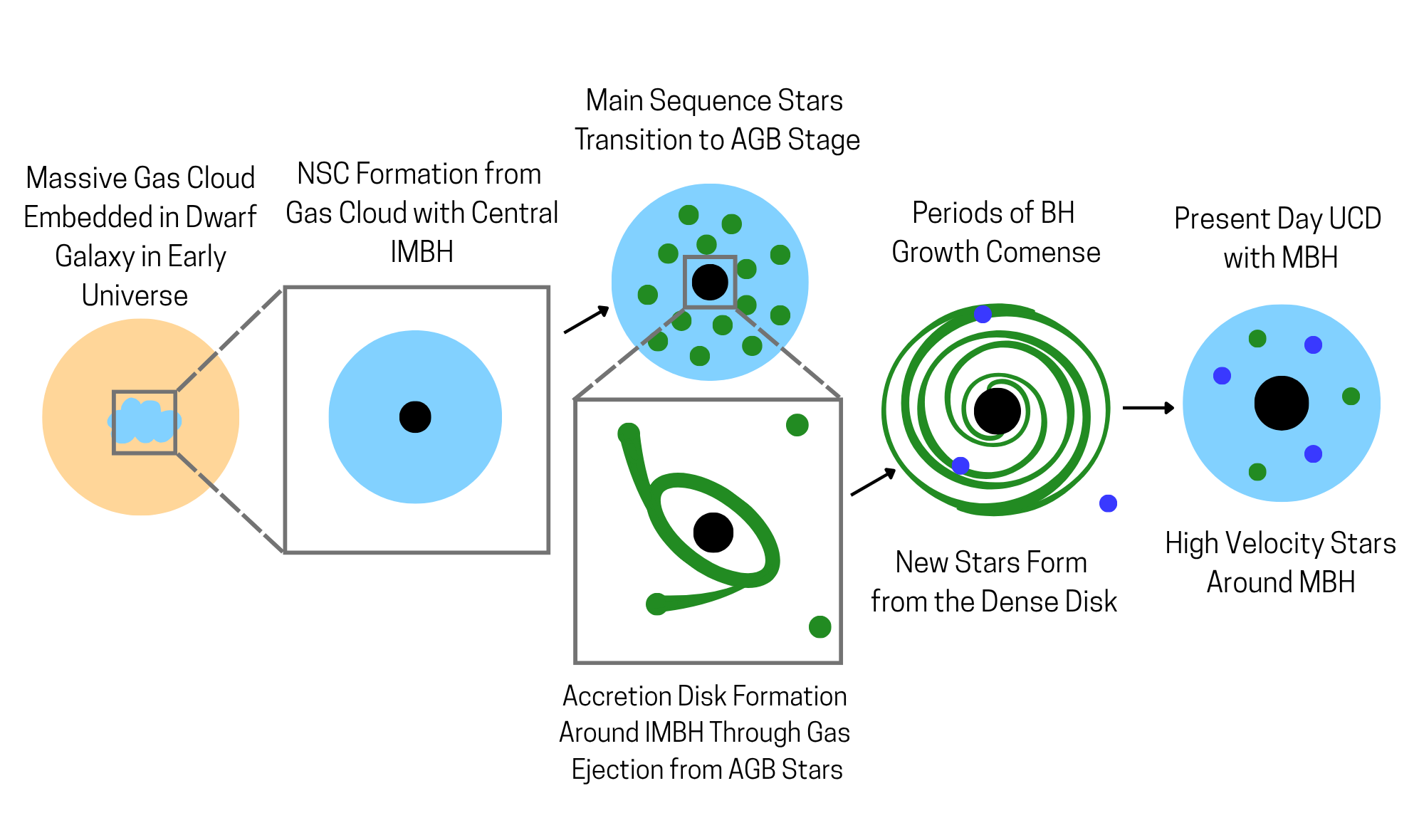} 
    \caption{Schematic of the proposed mechanism for accretion disk formation and BH growth. The illustration presents the evolution of a UCD from a massive gas cloud, to present day UCDs with MBHs. The main idea presented is the formation of AGB stars over time can undergo gas ejection which can infall towards an IMBH forming an accretion disk around it. The schematic also highlights key mechanisms present, such as the formation of new stars within the accretion disk that can populate the entire UCD. Leading to a key prediction for the proposed mechanism, being the creation of a new stellar population.}
    \label{fig:illustration}
\end{figure*}

Fig. \ref{fig:illustration} illustrates the physical model and key processes behind the proposed mechanism of accretion disk formation that we investigated. Here, we considered a UCD that formed in the early universe from a gas cloud within a dwarf galaxy with an IMBH formed in situ. \hl{\textbf{For the purposes of this study, the formation of the IMBH is not of great importance and is instead treated as the initial conditions for the model. The task of understanding the creation of the initial IMBH can be conducted in future work.}} \hl{\textbf{After the formation of the IMBH}}, the higher mass main sequence stars transition into AGB stars, the number of which is determined by the IMF of the system, which is computed in the following section. \hl{\textbf{Eventually, the AGB stars will begin having their gas ejected via stellar winds and that gas will undergo dynamics which are gravitationally influenced by the central IMBH.}} We predicted that \hl{\textbf{the}} gas ejecta will fall towards the IMBH and form an accretion disk, facilitating BH growth. \hl{\textbf{These periods of BH growth can occur several times over the lifetime of the UCD depending on the timing between AGB stars having their gas ejected.}} After \hl{\textbf{several}} periods of BH growth, the AGB ejecta in the accretion disk is converted into BH mass, resulting in an MBH that current observations are finding in UCDs. \hl{\textbf{The simulations used in this study will model one of these periods of BH growth, starting after the first stars have their gas ejected. The results of one period of BH growth will provide insight into the processes required for an initial IMBH to grow into the observed MBHs.}} 

We emphasise a key prediction for the model, \hl{\textbf{that}} being the creation of high velocity stars forming from the \hl{\textbf{condensing of}} gas ejecta within the accretion disk. These stars can be ejected from the disk, but still be bound to the UCD, \hl{\textbf{effectively}} creating a separate stellar population with a unique age, metallicity and kinematics than the first generation population. These new characteristics can be measured \hl{\textbf{in observational surveys}}, however, due to the size and distance for most UCDs the type of studies that can \hl{\textbf{verify these properties}} will be very difficult to conduct. Some studies have probed into the stellar populations of UCDs and found evidence of multiple through observation \citep{hilker2015stellar} and simulations of UCD formation \citep{bekki2015formation}. \hl{\textbf{A potential origin for the younger stellar population in these systems could be through the formation of new stars via the proposed mechanism.}}

\subsection{The IMF and the age-mass relation of stars}

\begin{figure}
    \centering
    \includegraphics[width=1\linewidth]{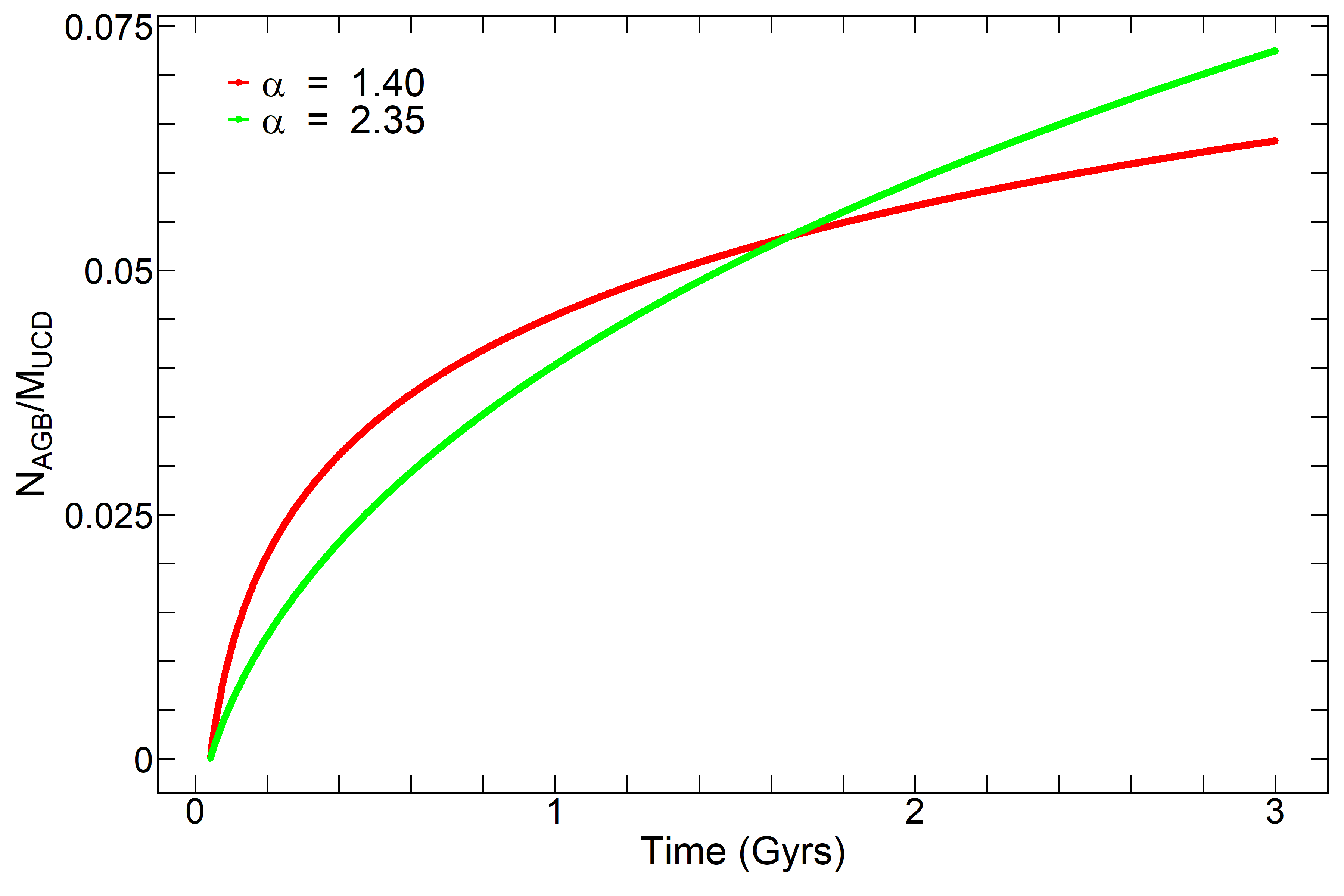}
    \caption{Cumulative number of AGB stars per mass of the \hl{\textbf{UCD}} over time for different IMF slopes. The red line represents a top-heavy IMF slope of $1.4$ and the \hl{\textbf{green}} line represents a standard IMF slope of $2.35$.}
    \label{fig:cumAGB}
\end{figure}


We used the IMF to calculate the number of AGB stars expected in a UCD of a given mass, \hl{\textbf{which is used in the fiducial model}}. The IMF power law below calculates the number of stars ($N_*$) within a particular mass range:

\begin{equation}
    \label{N_stars}
    N_{*} = \int_{m_{1}}^{m_{2}} C_{0} \ m^{-\alpha} \ dm
\end{equation}

Where \(C_{0}\) refers to the local stellar density and \(\alpha\) refers to the IMF slope. Solving for \(C_0\), we modify Eq. \eqref{N_stars} to solve for the stellar mass ($M_*$) contained within a particular mass range:

\begin{equation}
    \label{M_stars}
    M_{*} = \int_{m_{1}}^{m_{2}} C_{0} \ m^{1 -\alpha} \ dm
\end{equation}

Setting \(M_{*}\) to \(1\), we solve \(C_0\) for two IMF slopes. We use a top-heavy slope of \(\alpha = 1.4\) \citep{kroupa2002imf} and a standard slope of \(\alpha = 2.35\) \citep{salpeter1955luminosity}. Considering a mass range of \(0.1 - 50 M_{\odot}\), the two slopes result in \(C_0\) values of \(0.0588\) and \(0.1764\) respectively.

AGB stars have a mass range of approximately $0.5 - 8 M_{\odot}$. However, stars with a greater mass experience stages of their evolution at quicker rates and thus become AGB stars sooner. For our model, we only want to consider AGB stars that formed at a particular time after the initial SF period, \hl{\textbf{i.e. we only consider AGB stars that form at $t=t_0$}}. The lower mass bound of the AGB star population in a system decreases over time, which is visualised by the turning point in the Hertzsprung–Russell diagram (HR). This lower mass bound will be used to calculate the number of AGB stars at the start of the simulations. The mass value at the turn-off point on the HR diagram is described by the following relationship \citep{1986ASSL..122..195R} for mass in $M_{\odot}$ and time in years:

\begin{equation}
    \label{tmsp}
    \log_{10}[m] = 0.0434 \ (\log_{10}[t])^2 - 1.146 \ \log_{10}[t] + 7.119
\end{equation}

\begin{figure}
    \centering
    \includegraphics[width=1\linewidth]{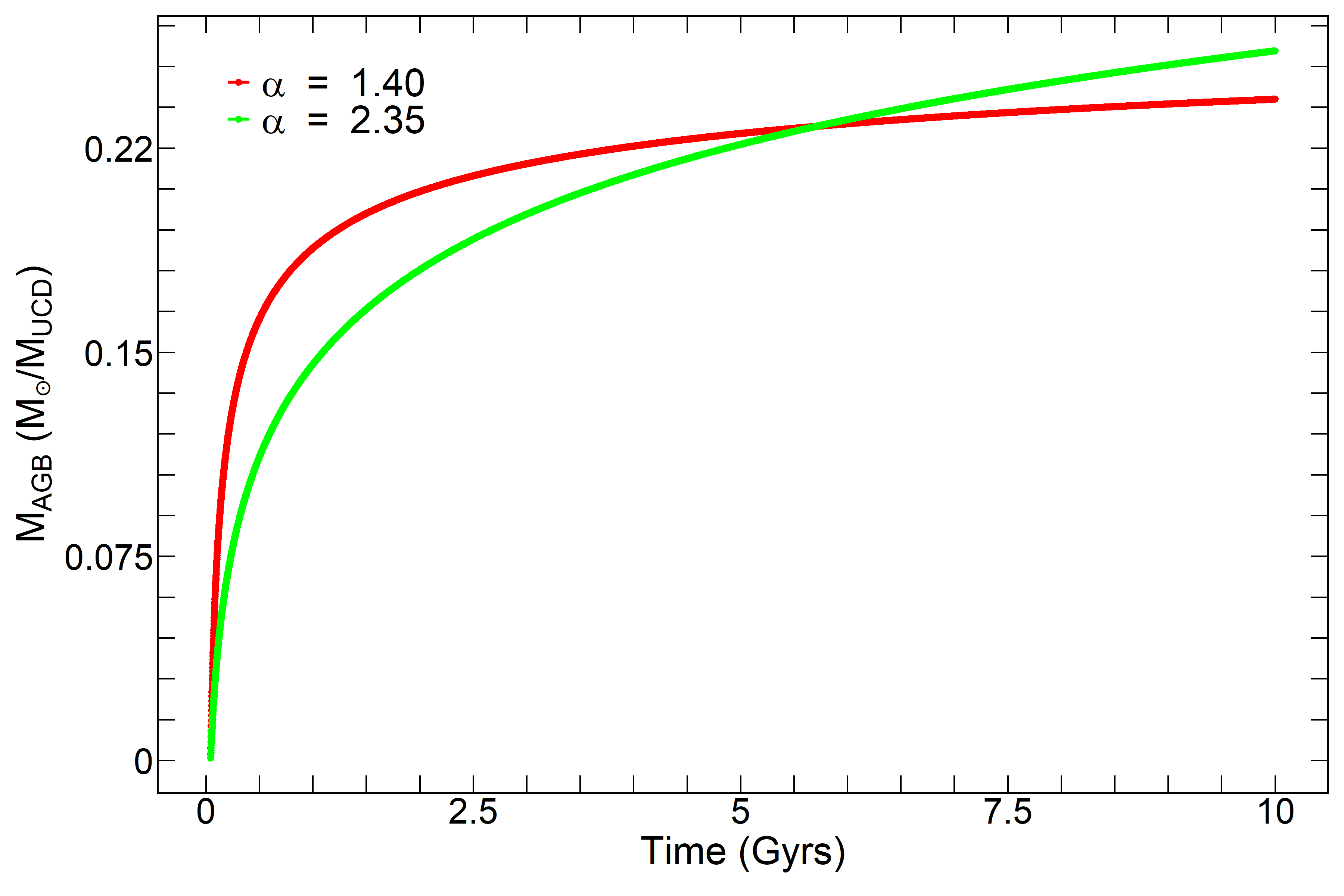}
    \caption{Cumulative mass from AGB stars per UCD mass over time for different IMF slopes. The red line represents a top-heavy IMF slope of $1.4$ and the \hl{\textbf{green}} line represents a standard IMF slope of $2.35$. The time axis was extended to \hl{\textbf{encapsulate}} typical ages of UCDs \hl{\textbf{for the purpose of determining the amount of AGB gas present for some calculations}}.}
    \label{fig:cumAGBmass1}
\end{figure}

Solving for the amount of time until $8\text{ M}_{\odot}$ stars reach the AGB stage, we consider AGB stars that form after $t\approx4.25\times10^7 \text{ yrs}$. \hl{\textbf{The motivation behind this choice for time is a computational one, that being if a lot of time passes there will be too many stars to model, requiring less particles per star to be used. However, if only a little amount of time passes then just a few stars are predicted. A few amount of stars are unsatisfactory, as we will explain in Section}} \ref{sec:physmech}, \hl{\textbf{simulations dedicated to only allowing a few set number of AGB stars to interact with the IMBH were conducted. Thus, this time value was chosen with the aim of producing more AGB stars than planned parameter dependence simulations but not too many AGB stars that the hydrodynamics of each star is not sacrificed.}} Using \hl{\textbf{the upper bound of $8\text{ M}_{\odot}$, and a lower bound determined by Eq. }}\eqref{tmsp} with Eq. \eqref{N_stars}, the number of AGB stars per \(M_{\odot}\) is \(\approx0.002236\) for \(\alpha = 1.4\) and \(\approx0.000869\) for \(\alpha = 2.35\). In our simulations, the default mass of the parent UCD is \(10^6 \text{ M}_{\odot}\). This leads to an estimation of the number of AGB stars of \(\approx2236\) and \(869\) for their respective IMF slope. Fig. \ref{fig:cumAGB} shows the cumulative number of AGB stars per $M_{\odot}$ over time \hl{\textbf{and the relationship between the number of AGB stars and IMF slope}} using the procedure described above. 

We can estimate the total AGB mass available for an accretion disk to form by recasting Fig. \ref{fig:cumAGB} with Eq. \eqref{M_stars} \hl{\textbf{to compute the cumulative gas mass available over time}}. Fig. \ref{fig:cumAGBmass1} shows the expected mass of \hl{\textbf{all}} AGB stars in a UCD for a given time. With a typical age of $\gtrsim10 \text{ Gyrs}$, a UCD could contain $22\%-25\%$ of its total mass in AGB stars depending on the IMF slope considered. This sets an upper limit for which an IMBH can grow via the proposed mechanism. 

In Section \ref{Discussion3}, we discuss the capture efficiency of AGB gas around the IMBH and found that $\approx60\%$ of the total AGB mass was bound within $1 \text{ pc}$ of the central BH for a large number of AGB stars. Together with estimation of the total available AGB mass \hl{\textbf{(Fig.}} \ref{fig:cumAGBmass1}) and properties of observed UCDs, we \hl{\textbf{have determined that some UCD systems}} could have formed \hl{\textbf{MHBs}} via the proposed mechanism. It is more useful to calculate the \hl{\textbf{smallest}} initial BH mass needed to form present day MBH masses:
\begin{equation}
    \label{eq:initialBH}
    M_{bh_i} = M_{bh_f} - \Delta M 
\end{equation}

Where $M_{BH_i}$ and $ M_{BH_f}$ are the initial and final BH masses, respectively. $\Delta M$ is the amount of growth the BH undergoes \hl{\textbf{as predicted by the proposed mechanism}}, which can be approximated via:
\begin{equation}
    \label{eq:growthamount}
    \Delta M = f_{cap}(1-\epsilon)M_{AGB}(t)
\end{equation}
Where $f_{cap}$ represents the fraction of AGB ejecta that is captured within $1.0 \text{ pc}$ of the central BH. $\epsilon$ is the radiative efficiency and $M_{AGB}(t)$ is the total mass locked up in AGB stars for a particular \hl{\textbf{time, as determined by Fig.}} \ref{fig:cumAGBmass1}. \hl{\textbf{The purpose of finding $M_{BH_i}$ in this way is to find how much mass a BH must start with to produce their inferred MBHs if they grew via the proposed mechanism.}}

Data from \cite{mieske2013central} was used to \hl{\textbf{find observed UCDs with a high predicted BH mass fraction, along with other properties such as the UCD mass. The UCD masses were calculated via the equation: $\text{M}_\text{dym}=\text{M}_\text{pop} \times \sigma^2_\text{obs}/\sigma^2_\text{pop}$, which utilises the assumption that mass follows light. The BH masses are inferred through the assumption that the higher observed $\text{M}/\text{L}$ is a result of the central BH potential. Through this, the median BH mass fraction is found to be $11\%$. Using the inferred BH masses and UCD masses, we can estimate whether these systems could have started with IMBHs and utilised the proposed mechanism for their current BH mass. For}} this inspection, we will use $10 \text{ Gyrs}$ for the ages of all the UCDs, as the slope of the curves in Fig. \ref{fig:cumAGBmass1} are small for $t>10 \text{ Gyrs}$ \hl{\textbf{and because this procedure just an approximation to see whether the proposed mechanism could explain these high BH masses}}. In Section \ref{sec:initialBHs}, we discuss the model for thin accretion disks and their radiative efficiency, \hl{\textbf{specifically choosing a high radiative efficiency ($\epsilon=42\%$) to minimise the amount of BH growth.}} This choice was made in part because there is no consensus on these BH spins \hl{\textbf{in observations}} and we wish to show the initial mass BHs in the worst case. For this reason, an IMF slope of $\alpha=1.4$ was chosen, corresponding to a total available AGB mass of $24.3\%$ \hl{\textbf{of the UCD's mass}}. A table of these initial BH calculations is shown in Table \ref{tb:initialbh} along with the UCDs \hl{\textbf{(and their masses)}} they are a part of. Only UCDs containing a BH mass fraction $> 5\%$ were considered, \hl{\textbf{as the systems with high BH mass fractions are of main interest in terms of BH growth and testing the effectiveness of the proposed mechanism}}. 
\begin{table}
\centering
\caption{Table of UCDs and their properties along with an estimated value for an initial IMBH mass. UCD data from \protect\cite{mieske2013central} was used \hl{\textbf{for these calculations, specifically they provided the UCD mass ($M_{\rm UCD}$) and BH mass fraction ($M_{\rm MBH}/M_{\rm UCD}$).}}}
\resizebox{\linewidth}{!}{%
\begin{tabular}{llll}
     Name & $M_{\rm UCD}$ ($\times 10^{6} {\rm M}_{\odot}$) & $M_{\rm MBH}/M_{\rm UCD}$ & $M_{\rm bh_i}$ ($\times 10^{6} {\rm M}_{\odot}$) \\
        M59cO & 160 & 0.31 & 36.1 \\
        S417 & 31 & 0.4 & 9.78 \\
        S928 & 20 & 0.4 & 6.31 \\
        UCD1 & 37 & 0.15 & 2.42 \\
        F9 & 17 & 0.21 & 2.13 \\
        F24 & 36 & 0.11 & 0.91 \\
        VUCD1 & 32 & 0.11 & 0.81 \\
        0320 & 4.9 & 0.13 & 0.22 \\
        HGHH92-C1 & 6.8 & 0.11 & 0.17 \\
        VHH81-C5 & 4.6 & 0.12 & 0.16 \\
        S314 & 9.0 & 0.1 & 0.14 \\
        0365 & 5.4 & 0.096 & 0.06 \\
        S490 & 15 & 0.088 & 0.05 \\
        F8 & 14 & 0.087 & 0.03 \\
        VUCD4 & 27 & 0.071 & -0.37 \\
        VUCD3 & 65 & 0.078 & -0.43 \\
        UCD5 & 20 & 0.061 & -0.47 \\
    \end{tabular}%
    }
    \label{tb:initialbh}
\end{table}

\hl{\textbf{The main result presented in Table}} \ref{tb:initialbh}, \hl{\textbf{is that there are many MBHs in these systems whose origins could have been an IMBH which underwent BH growth via the proposed mechanism. Using the typical upper limit of an IMBH ($10^5 \text{ M}_\odot$), 12 of the 17 UCDs presented in Table}} \ref{tb:initialbh} \hl{\textbf{meet this criteria. Some of which even produce negative masses, which physically represent the scenario that the system could have started with an IMBH of any mass and still produce the MBH we observe today. The UCDs with $M_{\rm bh_i}$ in the IMBH range represent the systems which could have started with an IMBH, but the minimum mass of that IMBH is bounded by the $M_{\rm bh_i}$ value. The last UCDs which have $M_{\rm bh_i}$ above the IMBH range represent the systems which either had to start with a BH larger than an IMBH or have utilised different processes of BH growth, possibly alongside the proposed mechanism. This table shows for a range of UCD masses and BH mass fractions that the proposed mechanism provides adequate AGB mass and leeway for these systems to have started with a central IMBH. A greater discussion on the meaning and consequences of these results is provided in Section}} \ref{sec:initialBHs}.

\subsection{Simulations of gas accretion onto IMBHs} \label{sec:sim}
\begin{table*}
\centering
\caption{Table of models used for investigating accretion disks around IMBHs. Parameters include UCD stellar mass ($M_*$), BH mass ($M_{bh}$), number of AGB stars ($N_{AGB}$), gas temperature and wind velocity. Parameters with a range of values are either averaged across or plotted against each-other. The choice of which is discussed throughout the paper.} 
\begin{tabular}{lllllllll}
Model ID 
& $M_{\rm *}$ ($\times 10^{6} {\rm M}_{\odot}$)
& $M_{\rm bh}$ ($\times 10^{6} {\rm M}_{\odot}$)
& $N_{AGB}$
& Gas Temp (K)
& Wind Vel (km/s)
& SF
& DM
& Comments \\ 
Mf1 & 1.0 & 0.0001 - 0.1 & 869 & 10, 100, 1000 & 10, 20 & yes & no & fiducial model\\
Mf2 & 1.0 & 0.0001 - 0.1 & 2236 & 10, 100, 1000 & 10, 20 & yes & no &  \\
Ma1 & 0.01 & 0.01, 0.1 & 1-300 &  10, 100, 1000 & 10, 20 & yes & no &  \\
Ma2 & 0.1 & 0.01, 0.1 & 1-300 &  10, 100, 1000 & 10, 20 & yes & no &  \\
Ma3 & 1.0 & 0.01, 0.1 & 1-300 &  10, 100, 1000 & 10, 20 & yes & no &  \\
Ma4 & 10.0 & 0.01 & 1-300 &  10, 100, 1000 & 10, 20 & yes & no &  \\
Md1 & 1.0 & 0.05 & 2236 & 100 & 10 & yes & yes/no &  also varied BH position\\
Ms1 & 1.0 & 0.01, 0.05, 0.1 & 2236 & 10, 100, 1000 & 10, 20 & yes/no & no & \\
Ms2 & 1.0 & 0.01, 0.05, 0.1 & 869 & 10, 100, 1000 & 10, 20 & yes/no & no &
\end{tabular}\label{tb:models}
\end{table*}

In order to investigate (i) the dynamical evolution
of UCDs and NSCs, (ii) orbital evolution of IMBHs within them
and (iii) gas dynamics of gaseous ejecta from AGBs
in a self-consistent manner, we adopted our code for direct N-body simulations from the evolution of GCs in dwarfs
\citep{bekki2016formation, wirth2020formation}. In the present simulations,
the dynamical friction of IMBHs within UCDs and NSCs is included, and the gas ejection from
AGB stars within them is also newly modelled.
We adopted the smooth particle hydro-dynamics (SPH) method to follow the hydrodynamical evolution of gaseous ejecta from AGB
stars within UCDs and NSCs with IMBHs.
Each IMBH is modelled as a collisionless particle, and its growth and feedback on ISM are not
modelled in the present study. SF from gaseous ejecta is also included,
because SF can possibly regulate the formation and evolution processes
of gaseous accretion disks around IMBHs.

\hl{\textbf{The simulations conducted in the present study differ from the state of the art high resolution galaxy evolution simulations in a few key aspects. The first of which is the scale, where galactic simulations model systems $>100 \text{ kpc}$. However, this study focuses on individual clusters on the scale of $10 \text{ pc}$, thus some processes and mechanisms are not of interest. For example, galaxy evolution simulations typically include magnetic fields and feedback effects. These differences lend the present simulations to compliment galaxy evolution simulations.}}

\hl{\textbf{In the present study, we do not explicitly model the magnetic field of the UCD, as direct evidence for their values is limited. Also, the implementation of a magnetic field is not necessary for an initial study into the formation of accretion disks. Feedback effects from the BH are omitted for the same reason. Future studies can investigate the impact of magnetic fields, feedback effects and more complex mechanisms. However, for now, the purpose of this study is to begin understanding the proposed mechanism, the results of which could be used to guide a more complex model.}}

\hl{\textbf{The magnetic fields in the most massive GCs have been studied more successfully than UCDs. The internal fields of 47 Tuc were constrained to $66 \pm 11 \text{ } \mu\text{G}$ through observed values of rotation measures (RM)}} \citep{abbate2020constraints}. \hl{\textbf{The magnetic field of $\omega \text{ Centauri}$ has not been directly found, however, RM values have been found to be $-18 \pm 8 \text{ rad/m}^2$}} \citep{dai2023timing} \hl{\textbf{which is the first step in finding its actual value. Therefore, there is motivation to implement a magnetic fields in UCDs for future studies of the proposed mechanism.}}

We adopt the assumption that an initial UCD or NSC
has a Plummer spherical density profile
\citep{10.1063/1.2811635}
with a total stellar mass ($M_{\rm UCD}$),
and a size ($R_{\rm UCD}$).
In a Plummer model,
the scale length ($a_{\rm UCD}$) of the system is determined by the formula:
\begin{equation}
a_{\rm UCD} = GM_{\rm UCD}/6{{\sigma}_{\rm UCD}}^{2} \;
\end{equation}
where G is the gravitational constant and
${\sigma}_{\rm UCD}$ is
the central
velocity dispersion of the nucleus.
Although UCDs and NSCs could have internal rotation to a varying degree,
we did not consider the initial angular momentum
of NSCs in the present study and assumed that the velocity dispersion
is isotropic.  We will investigate how
such internal rotation and anisotropic velocity dispersion
can influence the formation processes of accretion disks around IMBHs
in our future work.
Each UCD is represented by $N=5 \times 10^5$ collisionless particles and the softening length
is determined by $N$ and $R_{\rm UCD}$ such that
the gravitational softening length is the mean particle separation at $R=0.2R_{\rm UCD}$.

A small fraction of stars in a UCD can evolve into AGB stars depending on the adopted IMF.
We properly estimated the number of AGB stars and the mass of ejecta from these AGB stars
for each UCD model with a given $M_{\rm UCD}$ and IMF slope.
We assume that one AGB star parcel can eject $N_{\rm ej}$ particles with an \hl{\textbf{initial}} velocity
corresponding to the typical wind velocity of AGB stars ($v_{\rm wind}=10$ km/s). \textbftwo{The choice of $N_{\rm ej}$ also determines the mass of each gas particle in the model, as the total mass of an AGB star must be divided amongst $N_{\rm ej}$ particles. Typical choices of $N_{\rm ej}$ allowed for gas particles to vary in mass ranging from $10^{-4}$ and $10^{-2}$ $\rm M_\odot$.} Since the AGB wind velocity can be different between different AGB stars with different progenitor
stellar masses, we investigated models with different $v_{\rm wind}$. The total mass of
AGB ejecta ($M_{\rm ej}$)
from one AGB star is calculated from the adopt IMF and the progenitor stellar mass.
\hl{\textbf{Studies of AGB winds have shown that winds with an initial temperature of $\approx1000 \text{ K}$ can cool rapidly at $R=10^4R_*$ (Where $R_*$ is the stellar radius) during the early expansion phases into the ISM. \citep{cherchneff2006chemical}. The present simulation can not resolve the early evolution of stellar winds. Therefore, we assume the initial temperature of AGB ejecta to be a free parameter ($T_{\rm g}$) ranging from $10-1000 \text{ K}$.}}
The AGB ejecta is assumed to have a spherical distribution with a radius of $0.1$ pc.
A separate gravitational softening length for gas particles $\epsilon_{\rm g}$
is chosen so that the sub-pc scale
hydrodynamics of the AGB ejecta can be investigated. In most models, $\epsilon_{\rm g}$ was \hl{\textbf{found to be between the values of $0.01 - 0.0235 \text{ pc}$}}

\textbftwo{The above gas model utilises similar assumptions made by recent models of stellar winds, specifically, the implementation of a $10 K$ temperature floor (e.g. \cite{vlemmings2024molecular, ceulemans2026rotational}). The rationality behind this temperature floor is to prevent the gas temperature from becoming unrealistically low \citep{cordiner2009density}. The results of which can negatively affect other important processes, such as enhancing SF to an unrealistic level.}

\textbftwo{While other studies also use a $10 K$ gas temperature floor, it has been shown that this temperature can be reached shortly after AGB winds commence. The temperature of a gas envelope that evolves with radius has been implemented by \cite{van2018determining} as $T(r)=T_*\left(R_*/r\right)^\epsilon$. Where $T_*$ and $R_*$ represent the stellar temperature and radius, while $\epsilon$ is the characteristic power-law. \cite{maes2023sensitivity} use this equation with varying $T_*$, $R_*$ and $\epsilon$ to show that gas ejecta can be sufficiently cooled to the gas temperature floor after reaching a radius of $\approx10^{16}$ or $1000 \text{ AU}$. The reason that these temperatures can be reached is through CO rotational transitions, which is abundant in AGB ejects and can occur even at low temperatures \citep{ceulemans2026rotational}. This process cannot be modelled in the present galaxy-scale model, as their resolution requirements are too large to correctly model these short distances. Also, since the temperature of AGB winds can reach $10 K$ at such low distances relative to that of the current simulations, setting initial gas temperatures to $10K$ is a valid initial condition. Thus, we introduce a simple gas model with the assumptions above, such that less impactful processes (e.g. the explicit cooling and temperature range of AGB winds) are not considered, while the more important processes are kept (e.g. hydrodynamical interactions and SF).} 

\textbftwo{The aforementioned studies implement a higher temperature ceiling than what is described for the current model to $\approx 3000 K$, because their main focus is to study the kinematics, chemistry and behaviour of AGB winds. This involves modelling the start of an ejecting envelope, while for the current study, our main focus is on the formation of accretion disks around IMBHs which occurs after the AGB winds would have sufficiently cooled. Thus, we do not require modelling the start of an ejecting gas envelope or high gas temperatures reaching $3000 K$.}

\textbftwo{We emphasise that the present study is a preliminary look into the formation of gas disks around an IMBH. More sophisticated models can be used in future work to explicitly model the processes not considered. A discussion of how the current gas model and a different, slightly more sophisticated model, compare is presented in Appendix \ref{app:newmodel}.}

SF is modelled by converting a gas particle into a new stellar one if
the gas particle satisfies the following physical condition:
\begin{equation}
\rho_{\rm g} \ge  \rho_{\rm g, th},
\end{equation}
where $\rho_{\rm g}$ and $\rho_{\rm g, th}$ are the gas density of the SPH particle
and the adopted threshold gas density of SF, respectively.
We adopted $\rho_{\rm g, th}=10^{10}\text{ atom/cm}^3$ as an appropriate value,
because it corresponds
to the mass density of the first stellar cores \citep{Meyer1978}. Therefore, our simulations
can be quite different from our previous simulations of SF in star clusters depending only
on the Jeans conditions \citep{bekki2006primordial, bekki2009origin}. In order to investigate how
these SF processes can influence gas accretion onto IMBHs, we also ran many models
without SF.

The dynamical evolution of gas ejected from AGB stars in UCDs can possibly be influenced
by dark matter haloes of UCDs' host dwarf galaxies. However, 
we mainly investigated gas accretion onto IMBHs in UCDs without dark matter. Firstly because
gravitational potentials of UCDs themselves could be more important than dark matter potentials
for gas dynamics of AGB ejecta, and secondly because
we need to avoid introducing an extra set of model parameters related to dark matter potentials.
We therefore investigated only several models with dark matter potentials of UCDs' host dwarf galaxies
in the present study.
We adopt the density distribution of the NFW
halo \citep{Navarro_1996} derived from previous CDM simulations
in order to describe the initial density profile of the dark matter halo
in a dwarf galaxy with a NSC and an IMBH:
\begin{equation}
{\rho}(r)=\frac{\rho_{0}}{(r/r_{\rm s})(1+r/r_{\rm s})^2},
\end{equation}
Where $r$, $\rho_{0}$ and $r_{\rm s}$ are
the spherical radius, the characteristic density of a dark matter halo and the
scale
length of the halo, respectively.
The $c$-parameter ($c=r_{\rm vir}/r_{\rm s}$, where $r_{\rm vir}$ is the virial
radius of a dark matter halo) and $r_{\rm vir}$ are chosen appropriately
for a given dark halo mass ($M_{\rm dm}$)
by using the $c-M_{\rm h}$ relation for $z=0$
predicted by recent cosmological simulations
\citep{neto2007statistics}. For the adopted mass ranges of $M_{\rm dm}$,
we consider that $c=16$ is a quite reasonable value.
In the present study, we mainly investigated dwarf galaxies with 
$M_{\rm dm}$ 
$=10^{10} {\rm M}_{\odot}$
and $R_{\rm vir}=24.5$ kpc. 
The distance of the UCD from the dark matter centre ($R_{\rm dist}$) is assumed to be a free parameter ranging from 0.1 pc to 200 pc. \hl{\textbf{There was a computational necessity to place the UCD off centre from the DM centre due to overflow errors. This would have an impact on results which consider the DM halo.}}

The resolution of the simulations do not allow for direct modelling of the vicinity of the IMBH, as the simulations are designed to capture the entire UCD. \hl{\textbf{In doing so, we can resolve the disk itself, but the explicit inflow of gas into the BH can not be resolved. Thus, even at this scale it is difficult to also model the vicinity of the BH and thus explicitly study the conversion of accretion disk mass into BH growth.}} Therefore, \hl{\textbf{the simulated disks did not represent accretion disks that explicitly contribute to BH growth}}, but rather the precursor to them. These disks can eventually evolve into \hl{\textbf{smaller disks which will be able to funnel gas mass to the BH, the results of which is explored in Section}} \ref{Discussion}. For the purposes of this study, the transition from the precursor disk to \hl{\textbf{a smaller}} accretion disk was handled theoretically. An explicit modelling of the vicinity of the IMBH can be studied in future works which can utilise this work to determine initial parameters and behaviours. 

We mainly investigated models with only one IMBH in the system (denoted by $N_{\rm IMBH}$) to more clearly understand the formation process of an accretion disk around an IMBH within a UCD.
Each IMBH in an UCD was represented by a collisionless point-mass particle with
a mass ($M_{\rm IMBH}$), and its initial distance with respect to the UCD's centre
is assumed to be a free parameter denoted as $R_{\rm IMBH}$.
The formation of accretion disks around IMBHs could be influenced by $N_{\rm IMBH}$, $M_{\rm IMBH}$,
and $R_{\rm IMBH}$, however, we mainly focused on the dependence of the results on $M_{\rm IMBH}$ in
the present study: The parameter dependence on $N_{\rm IMBH}$ and $R_{\rm IMBH}$ is briefly
discussed and will be investigated in detail in our future studies.
We mainly investigated the models whose parameters are summarised in Table \ref{tb:models}.

\section{Results}\label{Discussion}

\subsection{Fiducial Model}\label{Discussion2}

\subsubsection{Dynamics of accretion disk formation}
\begin{figure*}
    \centering
    \includegraphics[width=\textwidth]{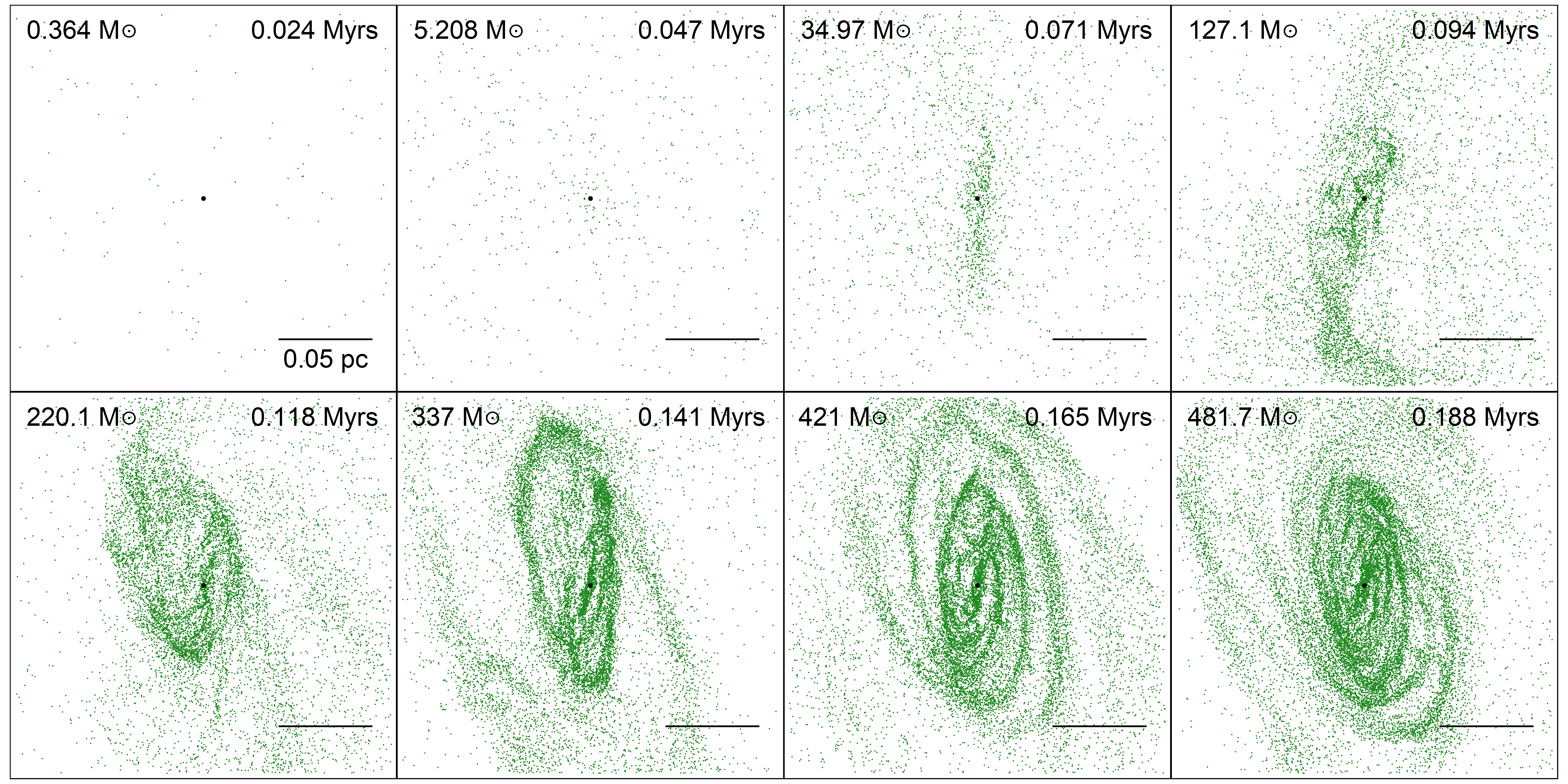} 
    \caption{Snapshots of the gas spatial distribution for the fiducial model (ID Mf1) for a short amount of time. The central black point represents the BH of mass $10^4 M_\odot$ ($1\%$ BH mass fraction) and the green points represent gas particles from AGB stars. The time of the snapshot is shown on the top right of each frame, similarly, the total gas mass contained within $0.1 \text{ pc}$ of the central BH is shown on the top left. Regarding the other parameters in the model used, the gas temperature was set to $10 \text{ K}$ and the gas velocity wind is initially $20 \text{ km/s}$.}
    \label{fig:best_gasdist}
\end{figure*}

\begin{figure*}
    \centering
    \includegraphics[width=\textwidth]{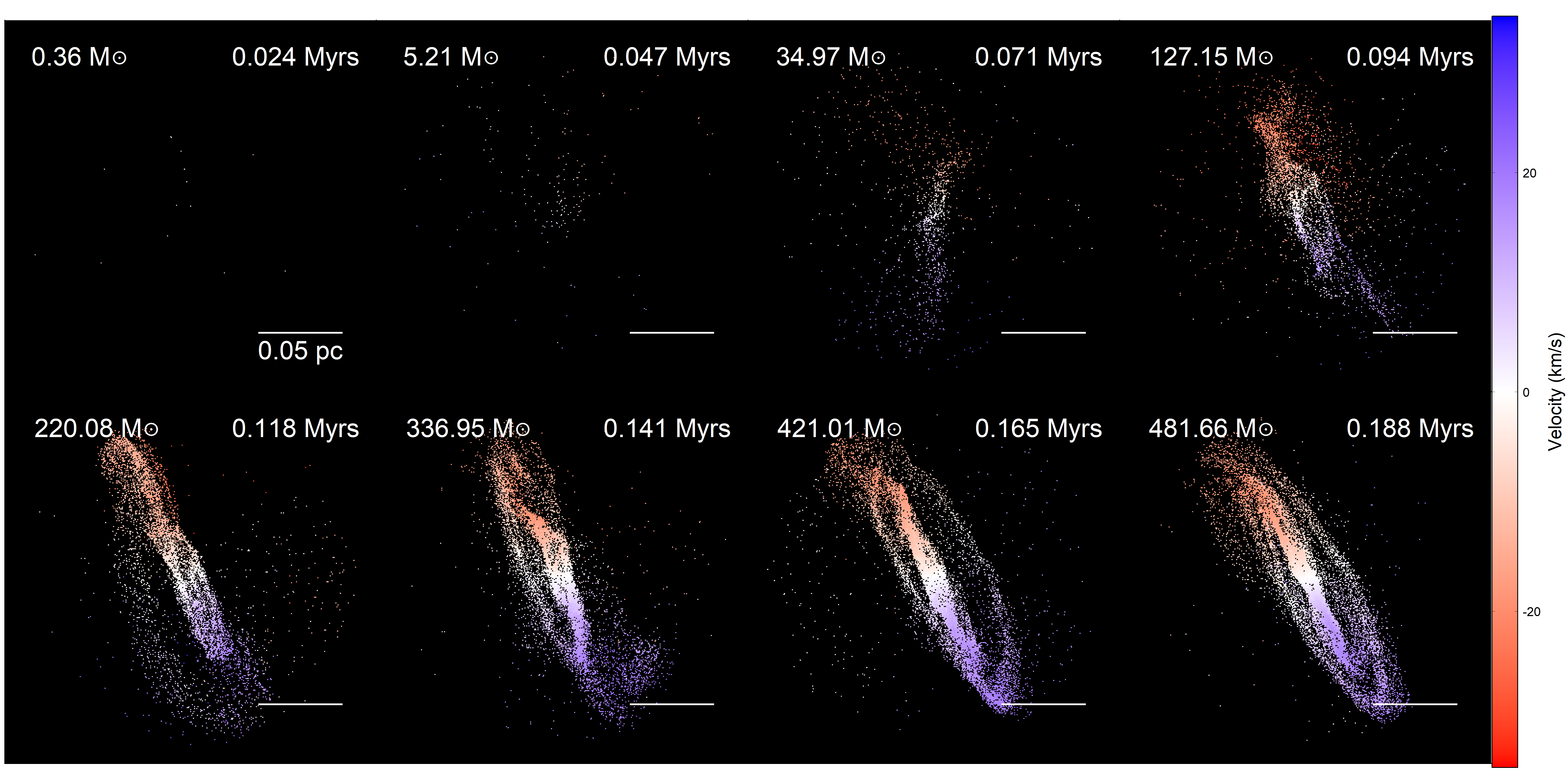} 
    \caption{Snapshots of the line of sight velocity distribution for the fiducial model, using the same simulation data and parameters as Fig. \ref{fig:best_gasdist} but from an edge-on viewing angle to best observe disk rotation. The centre of the snapshots coincide with the position of the BH, while the red and blue cells represent the mean line of sight velocity (km/s) of AGB gas particles within the individual cells. The magnitude and direction of the line of sight velocity is represented by the colour of the cell.}
    \label{fig:best_veldist}
\end{figure*}

Fig. \ref{fig:best_gasdist} displays snapshots of the distribution of the AGB gas (green points) over a short period of time to highlight the initial creation of the disk for the fiducial model. In this sequence of snapshots, \hl{\textbf{the amount of gas present within $0.1 \text{ pc}$ of the central BH increases over time. A large structure is observed to form at $t=0.071 \text{ Myrs}$ and continues to grow in both mass and size until the end of the simulation. These snapshots show the gradual accumulation of mass around the central BH, finalising at $481.7 \text{ M}_\odot$ after $0.188 \text{ Myrs}$. Throughout the snapshots, the structure of the disk exhibits more spirals as time progresses which are evident by visible arms within and extending out from the main mass. This is one indication that the observed structure is a spinning disk centred around a BH. To verify the characteristic rotation of the disk, Fig.}} \ref{fig:best_veldist} \hl{\textbf{shows the same simulation, instead with an edge-on view and measuring the line of sight velocity of the AGB gas particles. In these snapshots, the first observed net rotation is present at $t=0.071 \text{ Myrs}$ and continues until the end of the simulation. Verifying the characteristic rotation necessary for considering this structure as a rotating disk.}}

\begin{figure}
    \centering
    \includegraphics[width=\linewidth]{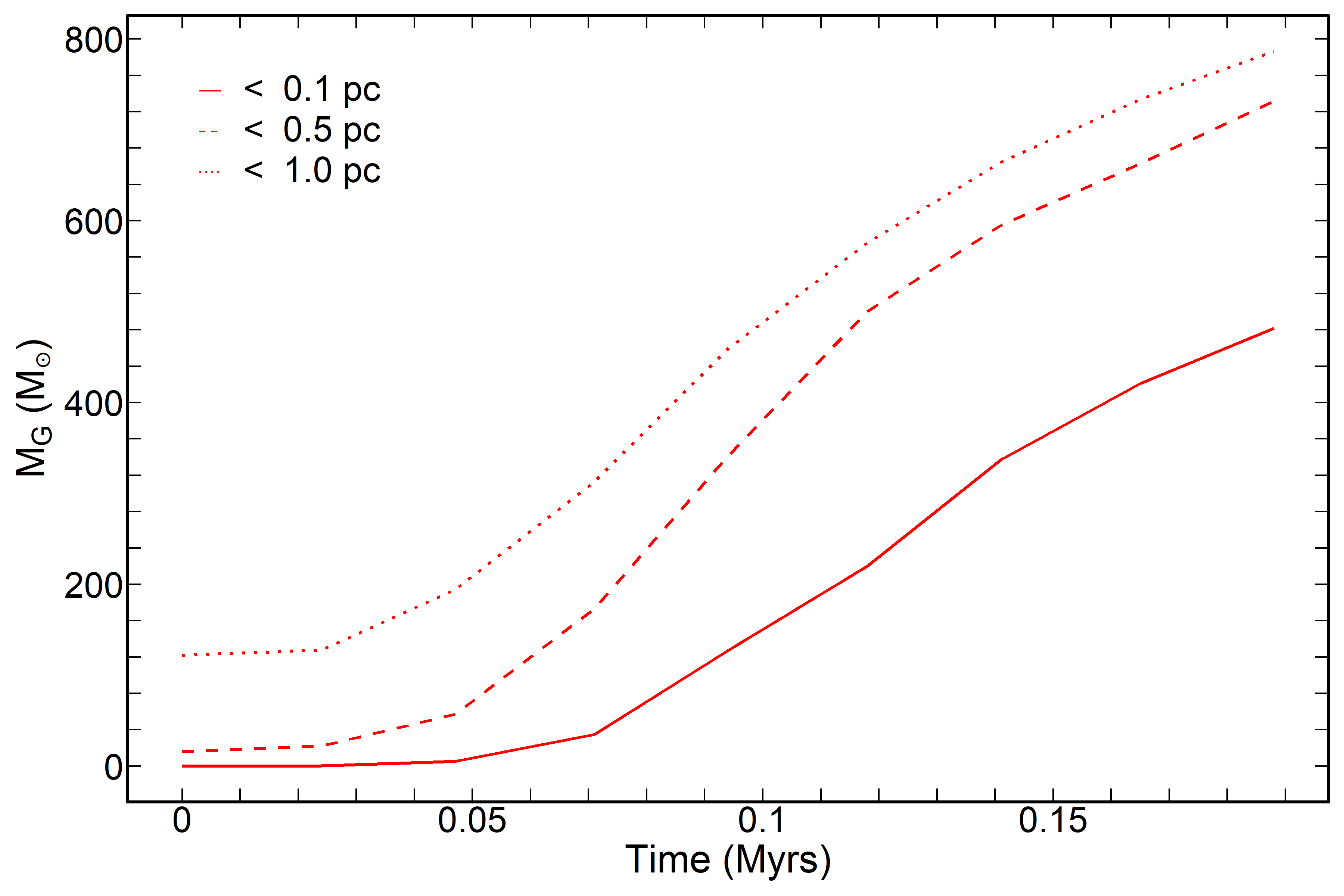} 
    \caption{Plot of the total gas mass found within $0.1 \text{ pc}$ (solid), $0.5 \text{ pc}$ (dashed) and $1.0 \text{ pc}$ (dotted) from the central BH over time in the fiducial model. The parameters used for this simulation are the same as the ones used to produce Fig. \ref{fig:best_gasdist}.}
    \label{fig:best_radialdist}
\end{figure}

\hl{\textbf{A more explicit view of the amount of AGB gas present near the central BH for the fiducial model is shown in Fig.}} \ref{fig:best_radialdist}, \hl{\textbf{which measures the total gas mass within $0.1 \text{, }0.5 \text{ and }1.0 \text{ pc}$. The main parameters used for the results presented are the same as those from Fig.}} \ref{fig:best_gasdist}. \hl{\textbf{Here, we observe not just the total amount of gas mass but also where that gas resides in the spiral. Initially, any gas that is found near the BH is found outside the $0.5 \text{ pc}$ mark, as a result of the gas not having enough time to fall near the BH. Over time, the total amount of gas near the BH increases, as expected from previous results and the initial gas distribution changes, showing a majority of the gas near the BH residing within the $0.5 \text{ pc}$ mark. This highlights the ability for gas in the outer parts of the main structure to migrate into the inner parts, resulting in a more massive and denser shape.}}

\hl{\textbf{The expected radius of an accretion disk is determined by the mass of the central object maintaining the disk. For a typical thin disk model, the radius of the disk is}} related by $R\propto M^{2/3}_{bh}$ \citep{shakura1973black}, with future research denoting a characteristic radius $R_{2500}$, corresponding to the peak emission of the $2500$ angstrom line. This characteristic radius can be related to the BH mass via \citep{morgan2010quasar}:
\begin{equation}
    \label{eq:bhrad}
    \log{\bigg\lvert\frac{R_{2500}}{\text{cm}}\bigg\lvert}=(15.78\pm0.12)+(0.8\pm0.17)\log{\bigg\lvert\frac{M_{bh}}{10^9 \text{M}_\odot}\bigg\lvert}
\end{equation}

\hl{\textbf{The above relation was derived via observations of 11 quasars which have a mass range of $\approx 10^8-10^{10} \text{ M}_\odot$}. This leads to the relation not covering the mass range we wish to consider ($10^4 \text{ M}_\odot$), however, the relation is consistent with the thin disk radius relation of $R\propto M^{2/3}_{bh}$ which is sufficient for the purposes of this analysis.} Using this equation, we estimate the expected radius for the accretion disk surrounding a $10^4 \text{ M}_\odot$ IMBH to be $R_{2500}\approx1.95\times10^{-7} \text{ pc}$, which is much smaller than what is simulated \hl{\textbf{by several orders of magnitude. This result highlights an important aspect of these simulations and their purpose in studying BH growth, specifically, these disks are precursors to the accretion disks which will enable BH growth. The main reason we did not model the vicinity of the BH and inquire about these accretion disks was due to the spacial resolution of the simulations. The small size of expected accretion disks was incompatible with the size required to observe the dynamics surrounding stars and the hydrodynamics of constituent gas particles, which is the main goal of the present study.}}

\begin{figure}
    \centering
    \includegraphics[width=\linewidth]{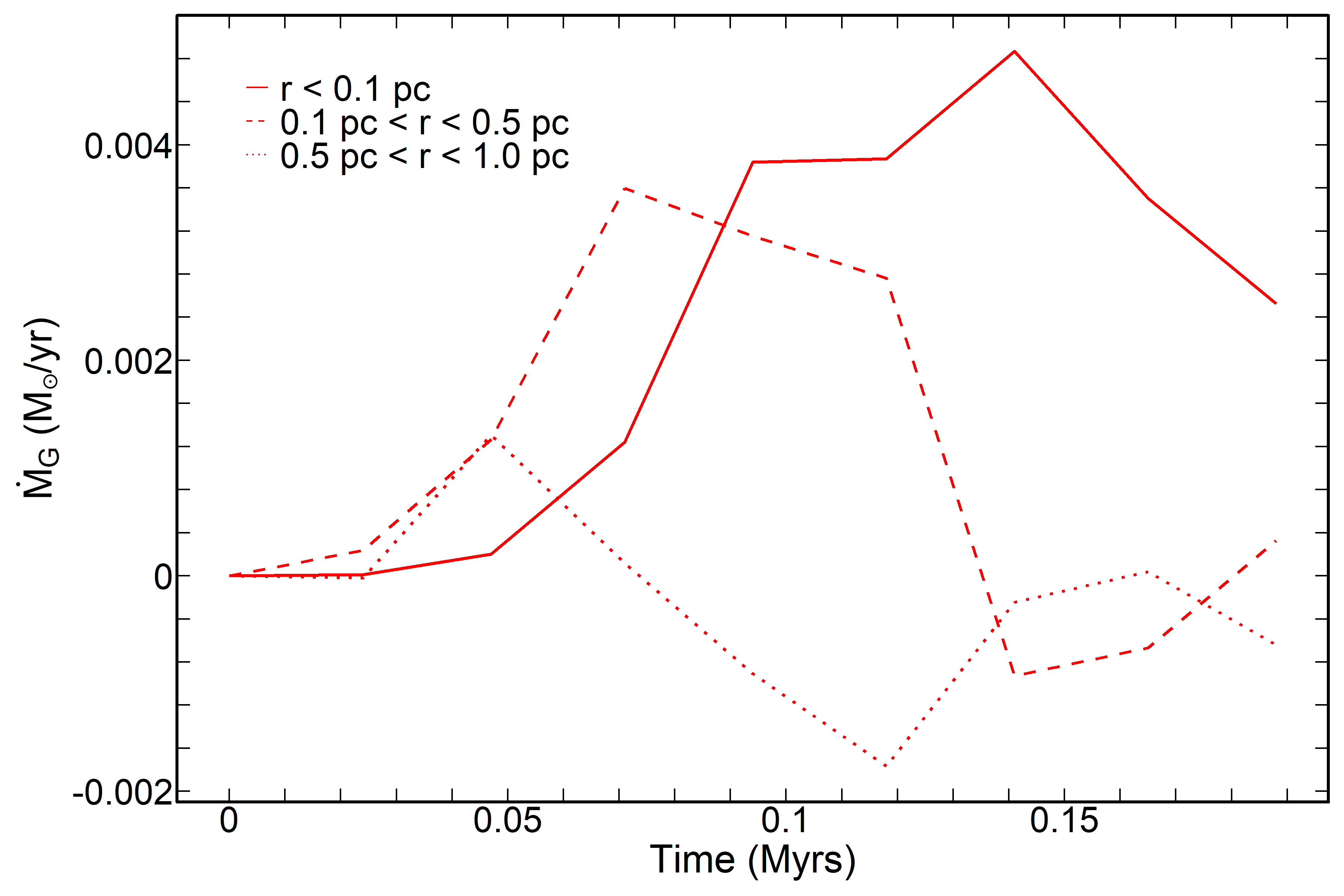} 
    \caption{The accretion rate of AGB gas onto different radius shells around a central BH over time. The outer shell contains the region of $1.0 - 0.5$ pc (dotted), the middle shell contains the region of $0.5 - 0.1$ pc (dashed) and the inner shell contains the region $< 0.1$ pc (solid). The parameters used for this simulation are the same as the ones used to produce Fig. \ref{fig:best_gasdist}.}
    \label{fig:best_accrdist}
\end{figure}

\hl{\textbf{Another important aspect of accretion disks is the accretion rate of matter. Fig.}} \ref{fig:best_accrdist} \hl{\textbf{measures the accretion rate of AGB gas for different radial shells around the central BH, specifically an outer, middle and inner shell. The key result from this analysis is the difference in accretion rates over the different shells. Comparing the outer and inner shell, we observe that there is a net negative accretion rate for a majority of the time for the outer shell, while the inner shell always accretes more mass over all time. The outer shell in these particular moments have more gas mass leaving its region than entering. Looking at both Fig.}} \ref{fig:best_radialdist} and \ref{fig:best_accrdist}, \hl{\textbf{the leaving gas mass preferentially moves towards the central BH, as seen by the increase in gas mass and positive accretion rate for the $<0.5$ pc region. This behaviour is similarly seen when considering the middle and inner shells, where the middle shell after some time exhibits a net negative accretion rate while the inner shell continues its positive accretion rate. The overall accretion rate is much lower than expected of other accretion disks, a possible reason for this is the larger than expected radius of the disk discussed above. Over time, as the radius of the disk decreases, the accretion rate for the vicinity of the BH will increase, possibly to values that are expected from observation and theory. These results highlight a key mechanism for the formation of accretion disks, that being the migration of gas from the outer part of the disk to the inner disk. In context with the previous discussion of the nature of these modelled disks, it provides confirmation that the simulated disks can eventually form accretion disks with small enough radii to begin BH growth via this gas migration mechanism. Strengthening the idea that the modelled disks here are progenitors of accretion disks and can contribute to BH growth after more gas migration. Here we will continue referring to these modelled disks as accretion disks, as the smaller accretion disks which directly contribute to BH growth and these larger disks which not as of yet contribute to BH growth are the same object.}}

\begin{figure*}
    \centering
    \includegraphics[width=\textwidth]{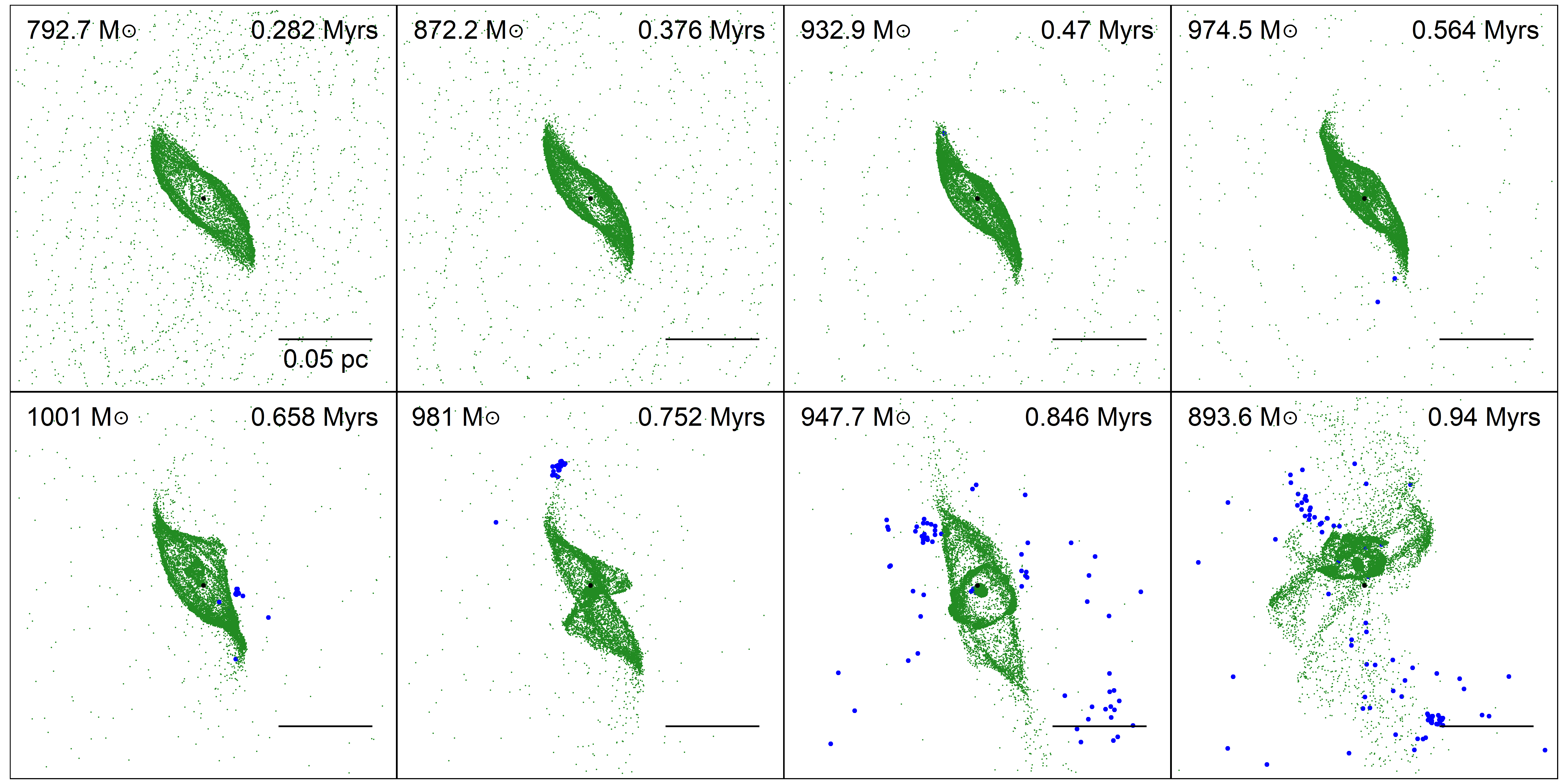} 
    \caption{Snapshots of the gas spatial distribution for the fiducial model (ID Mf1) over time. The central black point represents the BH of mass $5\times10^4 M_\odot$ ($5\%$ BH mass fraction) and the green points represent gas particles from AGB stars. The time of the snapshot is shown on the top right of each frame, similarly, the total gas mass contained within $0.1 \text{ pc}$ of the central BH is shown on the top left. In these snapshots, the conditions are such that new stars can form within the simulation time and are shown in blue. Regarding the other parameters in the model used, the gas temperature was set to $10 \text{ K}$ and the gas velocity wind is initially $20 \text{ km/s}$.}
    \label{fig:best_gasdist2}
\end{figure*}

\hl{\textbf{Another}} important mechanism \hl{\textbf{impacting}} accretion disk formation is the creation of new stars formed from the dense parts of the disk. This effect is shown in Fig. \ref{fig:best_gasdist2} which shows \hl{\textbf{snapshots of}} the distribution of gas (green) and new stars (blue) in the fiducial model. The snapshots skip ahead to a time where a \hl{\textbf{disk has formed}} with $\approx793M_\odot$ of gas within $0.1 \text{ pc}$ of the BH. Even at this time and mass, the amount of gas in the disk \hl{\textbf{still}} increases and the radius of the disk \hl{\textbf{decreases}}, leading to a more centrally dense disk. However, at around $0.658 \text{ Myrs}$, new stars start to form and the structure of the disk looks less symmetrical \hl{\textbf{and distorted}}. These new stars are present in the following snapshot, where the disk becomes even more disturbed. It is in this snapshot that, for the first time, the amount of gas present within $0.1 \text{ pc}$ decreases. The amount of gas removed from the disk is not only a consequence of gas escaping, but \hl{\textbf{also}} from the stars locking up the gas in stellar mass. The rest of the snapshots follow this trend, with more stellar particles present and the disk becoming less massive, less coherent and \hl{\textbf{even}} off centre, with the final mass of the accretion disk shown to be $\approx894 M_\odot$. These results highlight how SF can \hl{\textbf{act as a regulatory mechanism for the total mass and density of these disks, which in-turn regulates the amount of potential BH growth the system can undergo. SF in the disk can even lead to the destruction of disks, such as with the case shown in Fig.}} \ref{fig:best_gasdist2}. \hl{\textbf{All together, this places SF as a mechanism which works against the longevity of accretion disks, negatively affecting future BH growth.}}

\begin{figure}
    \centering
    \includegraphics[width=\linewidth]{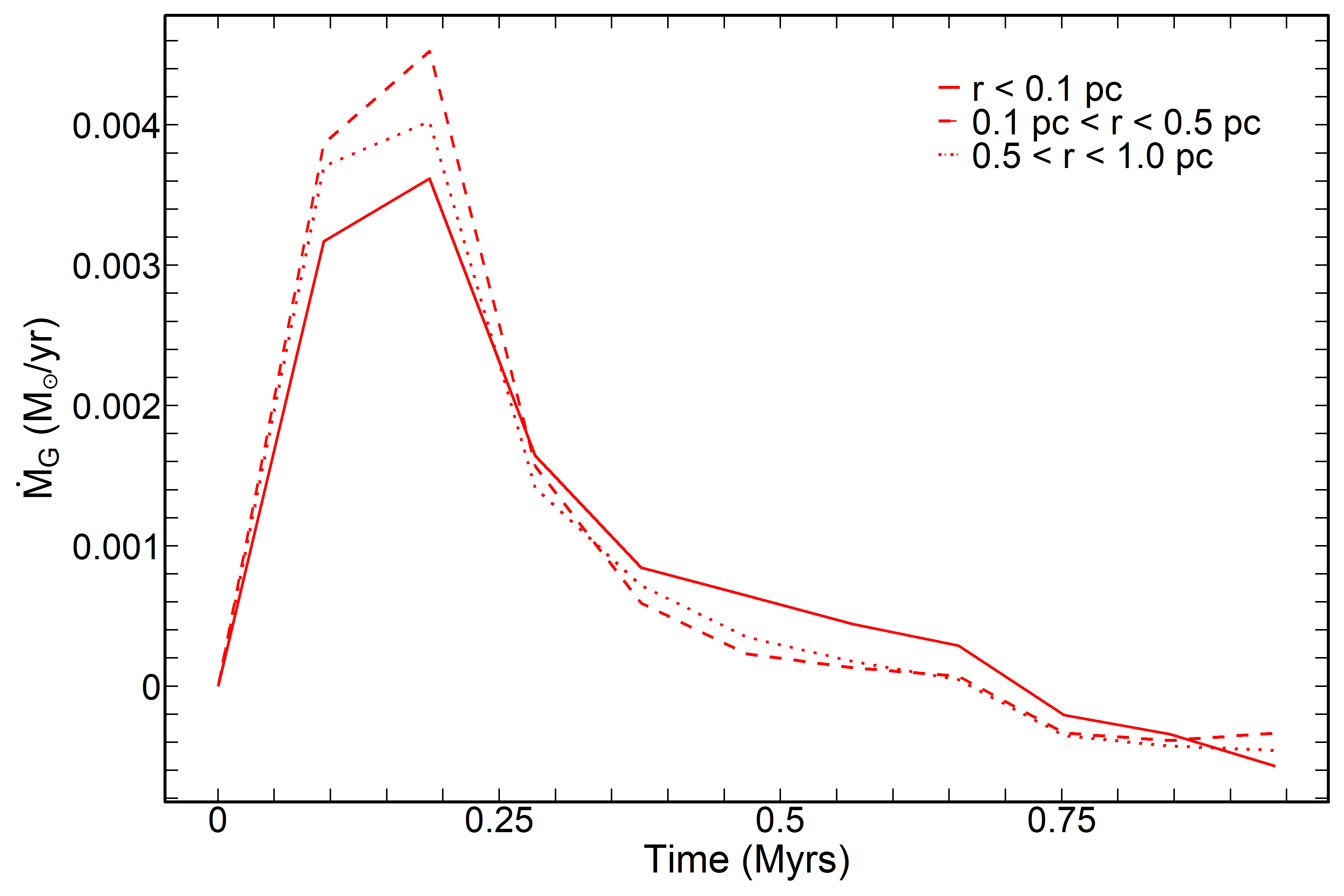} 
    \caption{The accretion rate of AGB gas onto different radius shells around a central BH over time. The outer shell contains the region of $1.0 - 0.5$ pc (dotted), the middle shell contains the region of $0.5 - 0.1$ pc (dashed) and the inner shell contains the region $< 0.1$ pc (solid). The parameters used for this simulation are the same as the ones used to produce Fig. \ref{fig:best_gasdist2}.}
    \label{fig:best_accdist2}
\end{figure}

\hl{\textbf{To highlight the regulatory effect of SF on the growth of the disk, Fig.}} \ref{fig:best_accdist2} \hl{\textbf{presents the accretion rate of gas mass onto different radial shells, similarly to Fig.}} \ref{fig:best_accrdist}. \hl{\textbf{However, the data used was from Fig.}} \ref{fig:best_gasdist2}, \hl{\textbf{which highlighted how the formation of new stars destroyed the disk structure. From the measurement of the accretion rate for the different shells, we see that accretion into the inner part of the disk is much larger than the previous figure showed. We also see that the accretion rate onto all the shells slows down after a short amount of time and finally reaches negative values at around $t=0.7 \text{ Myrs}$. The main differences between the behaviours of the two models was how the accretion rate evolved over time. Specifically, the disk that was destroyed by SF had a very high initial accretion rate, including the inner disk. While the first disk showed a slower accretion rate initially. Fig.}} \ref{fig:best_accdist2} \hl{\textbf{highlights how SF affects the growth of the disk in the form of slowing down the accretion rate and eventually reducing the rate to negative values for all the shells. Since all the shells exhibit the same decrease in the accretion rate, the loss of gas mass is not due to migration between shells but the removal of gas from the disk. The most obvious removal is through the conversion of gas mass to stellar mass, however the new stars that form also result in outward gas migration as highlighted in the later panels of Fig.}} \ref{fig:best_gasdist2}.

\begin{figure*}
    \centering
    \includegraphics[width=\textwidth]{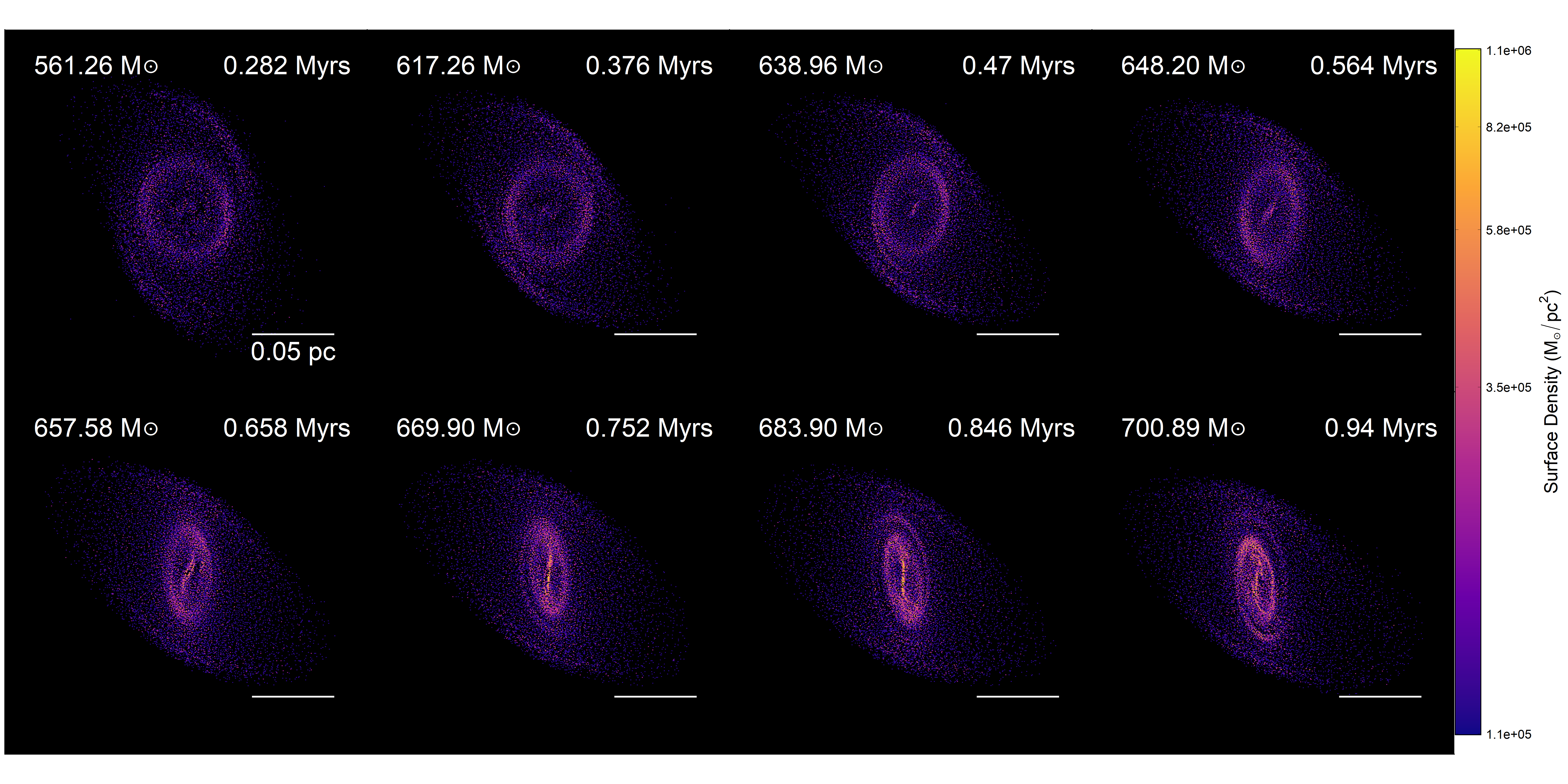}
    \caption{Surface density snapshots of the ejected mass from AGB stars measured in $\log|M_{\odot}/pc^2|$ over time. The centre of the plots lies a $10^4 M_\odot$ IMBH ($1\%$ mass fraction) and the simulation was run with a model ID of Mf1, specifically with a gas temperature of $10 \text{ K}$ and wind velocity of $20 \text{ km/s}$. The top left annotation for each frame represents the total mass contained within $0.1 \text{ pc}$ and the top right annotation denotes the time of the snapshot.}
    \label{fig:best_density}
\end{figure*}

\hl{\textbf{Fig.}} \ref{fig:best_density} \hl{\textbf{presents snapshots the surface density of the fiducial model over time. These snapshots aim to provide extra evidence for the mechanism of gas migration from the outer parts of the disk to the inner parts. As evident by the increased surface density present in the inner parts of the disk as time increases. In the beginning of the snapshots, the disk is relatively diffuse with most of the gas observed to reside in the outer parts. A circular structure is shown at an initial radius of $\approx 0.025 \text{ pc}$, and then decreases into the inner disk at a radius of $\approx0.01 \text{ pc}$. In the final snapshot ($t=0.94 \text{ Myrs}$), A total of $700.89 \text{ M}_\odot$ of AGB gas is found within $0.1 \text{ pc}$ of the central BH, with most of that mass concentrated in a persistently decreasing disk of radius $\approx0.01 \text{ pc}$.}}

\subsubsection{Physical mechanism} \label{sec:physmech}
In order to understand the physical mechanism of accretion disk formation, we ran models with a set number of AGB stars ranging from 1 - 300, as described by model ID Ma3. These simulations were utilised to \hl{\textbf{study the}} dynamical and hydrodynamical interactions between \hl{\textbf{ejected gas particles of}} AGB \hl{\textbf{stars}} and how they can influence the formation of accretion disks. Higher resolution simulations of model Ma3 were conducted \hl{\textbf{for this purpose}}, specifically with AGB numbers of $1$, $2$ and $5$. \hl{\textbf{In this case, higher resolution simulations represent those which an increase in the number of constituent AGB gas particles per AGB star, effectively increasing the resolution of hydrodynamical interactions. The number of constituent particles was increased from $\approx30$ to $1500$ particles per star.}}

\begin{figure*}
    \centering
    \includegraphics[width=1\linewidth]{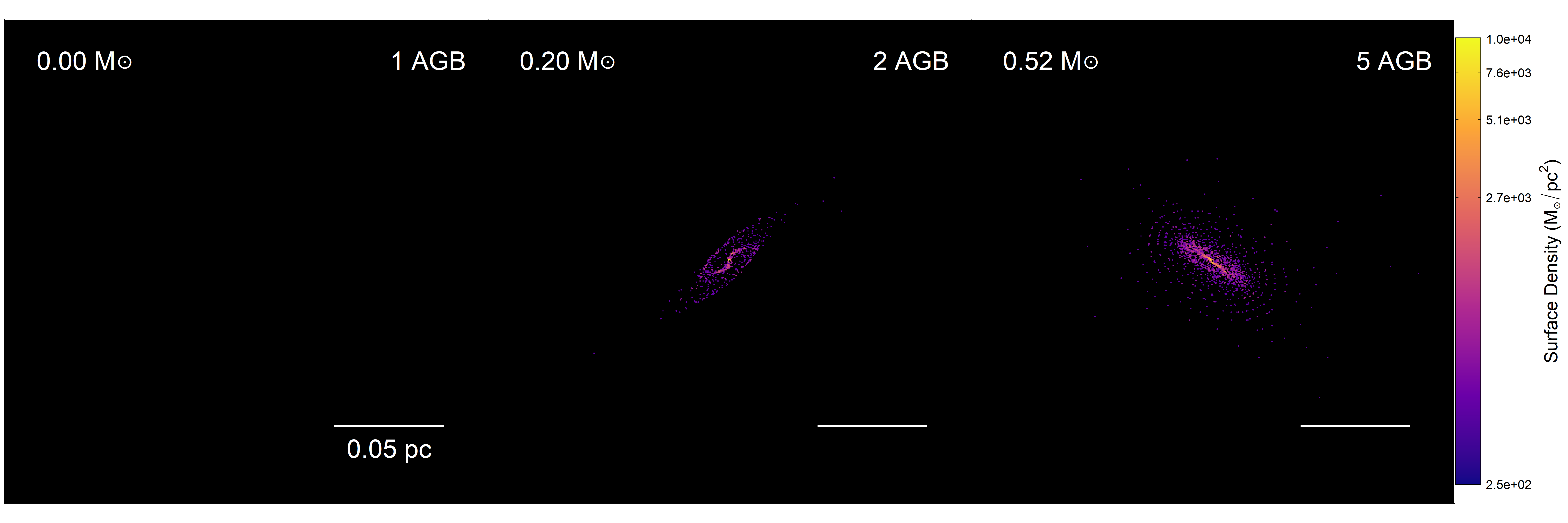}
    \caption{Surface densities of the final snapshots ($t=0.94 \text{ Myrs}$) from model Ma3 with a set number of AGB stars, specifically, $1$ (left), $2$ (middle) and $5$ (right). The parameters used for these simulations involved a UCD mass of $10^6 M_\odot$, a $1\%$ BH ($10^4 M_\odot$) with a gas temperature of $100 \text{ K}$ and wind velocity of $10 \text{ km/s}$. Annotated on the top left of each snapshot is the total gas mass within $0.1 \text{ pc}$ of the central BH.}
    \label{fig:agbdep:finaltime}
\end{figure*}

\hl{\textbf{Fig.}} \ref{fig:agbdep:finaltime} \hl{\textbf{presents the final snapshots of the surface density for model Ma3 with $1$, $2$ and $5$ AGB stars. For the $1$ AGB star case, after $0.94 \text{ Myrs}$, there is no AGB gas found within $0.1 \text{ pc}$ of the central BH. However, after increasing this to $2$ and $5$ AGB stars, a disk is able to form. The two low mass disks contain $\approx7.14\% \text{ and } 7.43\%$ of the total gas mass available from their initial AGB stars, respectively (the total gas mass available for the $2$ and $5$ AGB cases were $2.8 \text{ M}_\odot$ and $7 \text{ M}_\odot$, respectively). On top of the increase in the proportion of gas that is found in the disk, the $5$ AGB stars case is presented by a fuller and more centrally dense disk than the $2$ AGB case.}}

\begin{figure*}
\centering
\includegraphics[width=\textwidth]{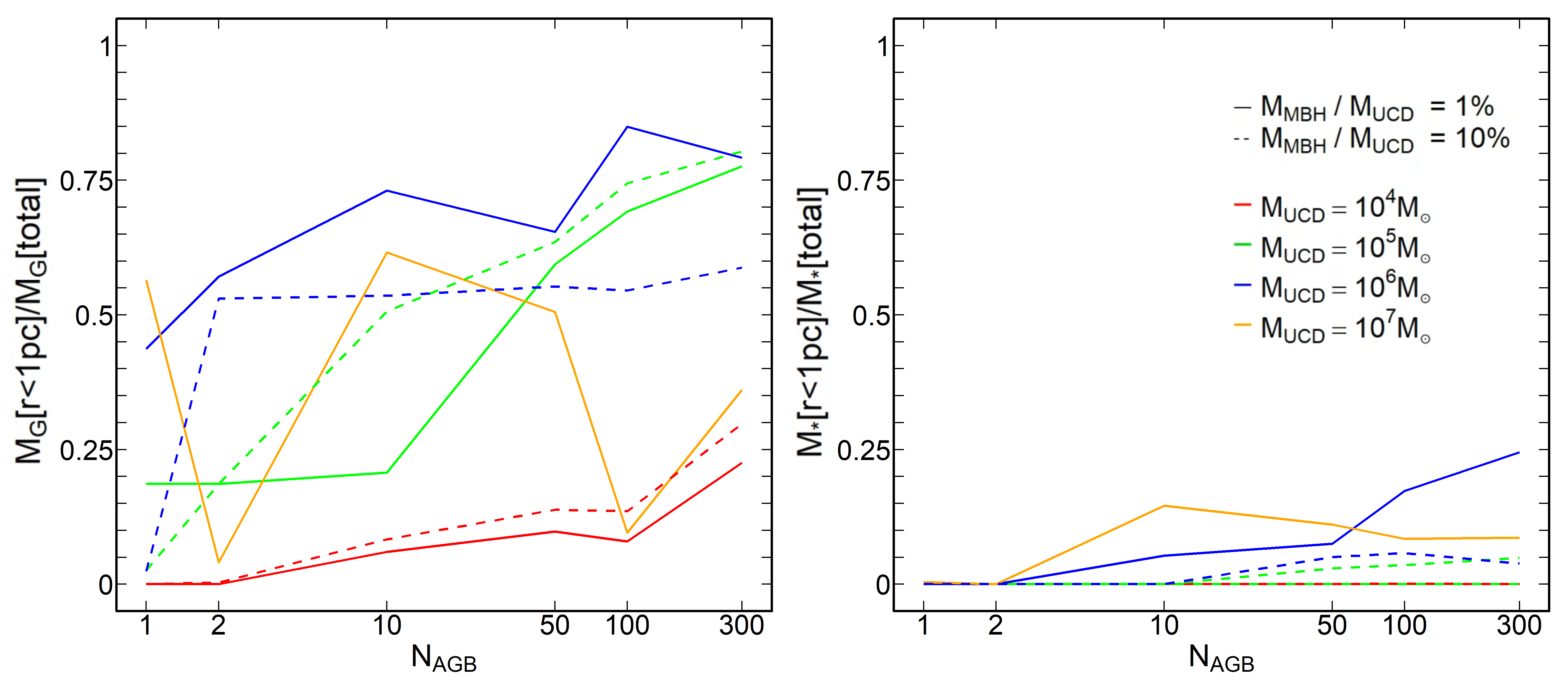}
\caption{AGB gas mass fraction \hl{\textbf{in terms of gas mass within $1.0 \text{ pc}$ of a central BH ($\text{M}_{\rm G}[r<1\text{pc}]$) over the total initial gas mass in the system ($\text{M}_{\rm G}[\text{total}]$)}} (left) and new star mass fraction \hl{\textbf{in terms of stellar mass within $1.0 \text{ pc}$ of a central BH ($\text{M}_{\rm *}[r<1\text{pc}]$) over the total mass of new stars ($\text{M}_{\rm *}[\text{total}]$)}} (right). \hl{\textbf{The results were computed from models Ma1 (red), Ma2 (green), Ma3, (blue) and Ma4 (orange) over a set number of AGB stars, including $1$, $2$, $10$, $50$, $100$ and $300$. The data is also grouped by BH mass fractions of $1\%$ (solid) and $10\%$ (dashed). These simulations ran for $\approx5 \text{ Myrs}$.}}}
\label{fig:agbdep:main}
\end{figure*}

Fig. \ref{fig:agbdep:main} plots the percentage of AGB gas mass and new stars within $1.0 \text{ pc}$ of the central BH for a varying number of AGB stars. Since the number of AGB stars affects the total available gas mass, bias in favour of simulations with a higher number of AGB stars is present. This bias is also present in the formation of new stars, as the maximum number of stars that can form depends on the available gas. To remove this bias, the analysis involved comparing the fraction of gas/stars present, \hl{\textbf{rather than the absolute amount}}. The total gas is determined by the amount of gas present at $t=0$. Only the initial total gas is used to normalise the results, \hl{\textbf{allowing for the effects of SF on gas loss to be measured.}} However, with SF increasing the total number of new stars over time and \hl{\textbf{there being no new}} stars at $t=0$, the \hl{\textbf{new star}} results were scaled by the total number of new stars in \hl{\textbf{a given}} time step. \hl{\textbf{The simulations conducted, were ran for an equivalent of $\approx5 \text{ Myrs}$ to allow enough time for a disk to form and potentially dismantle.}}

\hl{\textbf{Fig.}} \ref{fig:agbdep:main} \hl{\textbf{shows, for most of the parameter space, an increasing trend for the central gas mass percentage as the number of AGB stars increase.}} The existence of \hl{\textbf{this increasing trend is not due to presence of more gas, as explained by the analysis process above. Instead, the presence of gas from different origins results in a higher central disk density.}} This trend is most obvious for the $M_{UCD}=10^5 M_{\odot}$ and $10^6 M_{\odot}$ models, which \hl{\textbf{found for a large amount of AGB stars}} a final capture percentage of $\approx60\%-75\%$. Other cases \hl{\textbf{did not show such a high central density, even for a high number of AGB stars. For example, the $M_{UCD}=10^4 M_{\odot}$ model for a large number of AGB stars was only capable of capturing $\approx25\%$ of the total gas in the system. However, the increasing trend of central density percentage is still observed, even though it is smaller than the other models. The only model in the parameter space which do not follow this trend are the $M_{UCD}=10^7 M_{\odot}$ simulations, as it is shown to vary greatly in capture efficiency with no dependence on the number of AGB stars present. A greater discussion into why this scatter occurs is conducted in Section}} \ref{Discussion3}. \hl{\textbf{Besides the very high UCD mass simulations, the presence of more AGB stars results in a higher percentage of the total gas to be centrally located. Implying that interactions between gas ejecta particles from different AGB stars are important for those particles to infall towards the central BH. Results from both Fig.}} \ref{fig:agbdep:finaltime} and Fig. \ref{fig:agbdep:main} \hl{\textbf{supports the premise that interactions between different AGB stars can have an effect on the mass of the disk. This hydrodynamical interaction becomes another important mechanism for accretion disk formation, as it allows for the gas particles to lose their kinetic energy and migrate closer to the central BH.}}

\begin{figure}
\centering
\includegraphics[width=\linewidth]{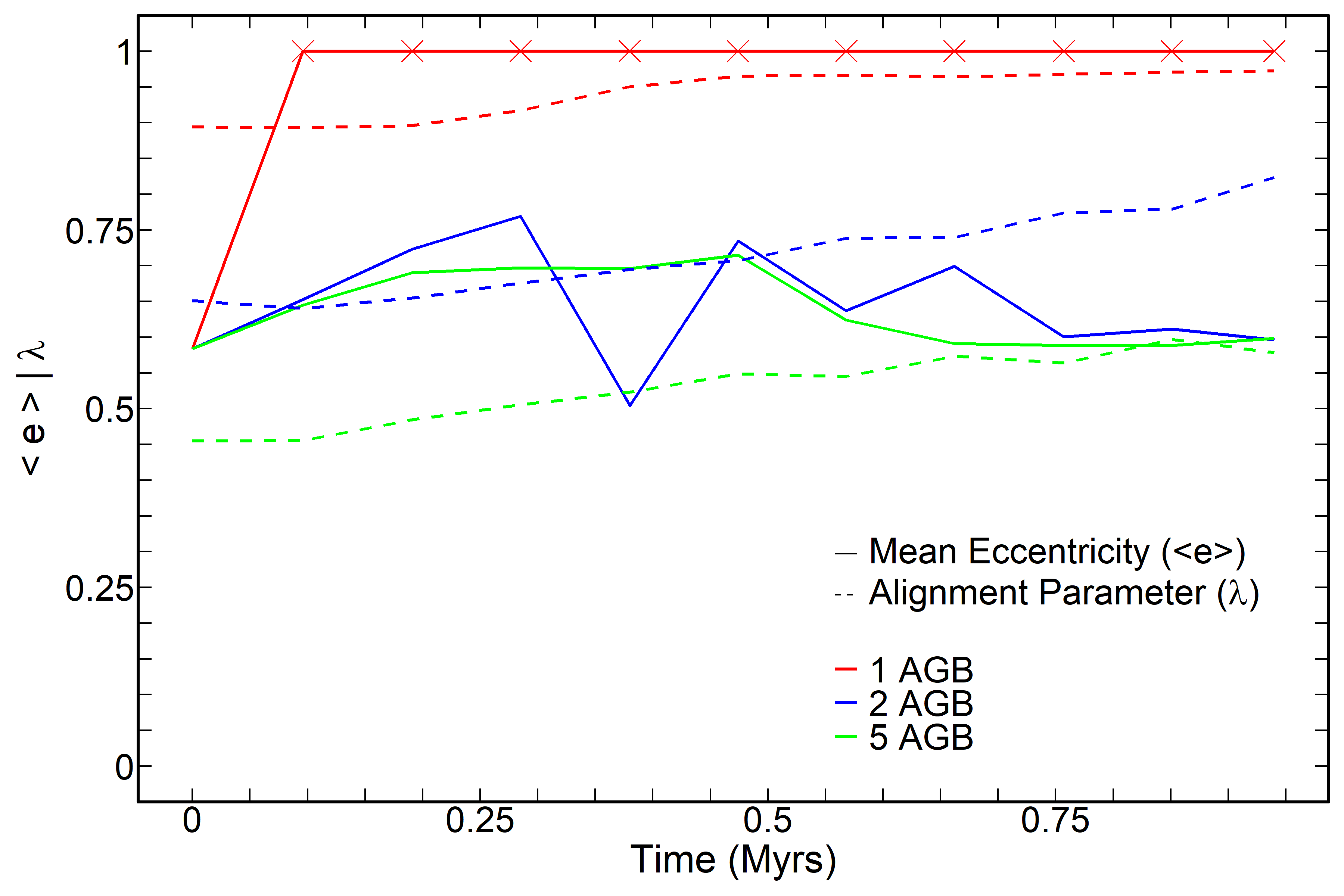}
\caption{The mean eccentricity ($\langle e \rangle$) (solid) and alignment parameter ($\lambda$) (dashed) over time for the same higher resolution models presented in Fig. \ref{fig:agbdep:finaltime}. The $1 \text{ AGB}$ (red), $2 \text{ AGB}$ (blue) and $5 \text{ AGB}$ (green) cases were considered. For the eccentricity calculation, only bound particles were considered. The crosses represent times where there were no AGB particles bound to the central IMBH.}
\label{fig:el}
\end{figure}

\hl{\textbf{We can verify that the interaction taking place is hydrodynamical in nature in Fig.}} \ref{fig:el}, \hl{\textbf{which plots the evolution of the mean eccentricity and alignment parameter. The alignment parameter is described by:}}

\begin{equation}
    \lambda = \frac{|\sum_i{m_i\vec{j}_i}|}{\sum_i{m_i|\vec{j}_i}|}
\end{equation}

\hl{\textbf{Where $\vec{j}_i$ represents the specific angular momentum of the ith particle and $m_i$ is the mass of the ith particle. This parameter describes how coherent the angular motion of the particles are and the value is normalised such that $0<\lambda<1$. Where 0 represents no coherent motion and 1 represents complete coherent motion. For a disk, we expect a high coherent motion.}} 

Fig. \ref{fig:el} \hl{\textbf{shows that all models start with high coherent motions and steadily increase over time. The initial high alignment for all models is due to the geometry of the initial conditions. Specifically, at the start, all the particles are moving with their ejection velocity from individual "sources", where the number of sources determine how coherent the rotational motion starts with. The more interesting result from this analysis is the relative increase in $\lambda$ being $8.8\%$, $26.5\%$ and $27.2\%$ for $1 \text{ AGB}$, $2 \text{ AGB}$ and $5 \text{ AGB}$, respectively. The two models which form disks from Fig.}} \ref{fig:agbdep:finaltime} \hl{\textbf{show a large relative increase in $\lambda$, implying that the presence of multiple AGB stars can more easily convert random motion into more coherent rotational motion. However, his alone is not enough to show that the interactions which create more coherent motion is hydrodynamical.}}

\hl{\textbf{The mean eccentricity was calculated by considering only particles which were bound, ensuring that the only possible values were between $0-1$. As a result, the $1 \text{ AGB}$ case immediately show NA values for the eccentricity because all the gas particles are not bound to the central IMBH. This result verifies that no disk can be formed, even though $\lambda$ increases (coherent motion outwards). The opposite is shown in the other 2 cases, where the mean eccentricity decreases from the initial value. We expect the eccentricity for a forming disk to decrease and circularise, which is exactly what is happening in the $2 \text{ AGB}$ and $5 \text{ AGB}$ cases. The combination of the eccentricity and alignment parameter results imply that the interactions are dissipating energy to both circularise and form coherent rotational motion, which is a hydrodynamical interaction.}}

\hl{\textbf{The right panel of Fig.}} \ref{fig:agbdep:main} \hl{\textbf{measures the percentage of the total new stellar mass present within $1.0 \text{ pc}$ of the BH. The main result of this plot is that all throughout the parameter space most of the new stars are not observed close to the BH, rather they are either ejected from the disk or maintain elliptical orbits. A discussion on some of the elliptical orbits is conducted in Section}} \ref{sec:omegacent}. \hl{\textbf{Those new stars which now reside in the UCD but not bound to the disk become part of a new stellar population. This new stellar population could be measurable, as this stellar population would be distinct from the old stars in terms of metallicity, age and kinematics. However, one of the main difficulties with studying UCDs observationally is their small size and large distances. A more thorough discussion on the new stellar population, how it can be used to verify the model and possible shared formation pathways with other systems is conducted in Section}} \ref{sec:omegacent}. 

\subsection{Parameter Dependence} \label{sec:paramdep}
\subsubsection{UCD/NSC mass}\label{Discussion3}
Fig. \ref{fig:agbdep:main} describes how the UCD mass affects the formation of accretion disks for different BH mass fractions. In the first panel, the fraction of gas trapped within $1 \text { pc}$ of the BH increases with increasing UCD mass, \hl{\textbf{except for the high mass case}}. This result is expected because with increasing the UCD mass the gravitational potential deepens \hl{\textbf{in the centre where the BH resides}}, allowing more gas to be gravitationally bound near the BH. \hl{\textbf{There is an expectation that this trend should follow for an increase in BH mass, as this would also increase the gradational potential at the centre. However, there are some examples in plot which show that this is not the case. Specifically, the $M_{UCD}=10^6 M_{\odot}$ (blue) results show that the smaller BH mass can capture a higher fraction of the total gas present. Also, for a large number of AGB stars, the smaller BHs in the $M_{UCD}=10^5 M_{\odot}$ (green) systems capture more gas than the $10\%$ BH in the $M_{UCD}=10^6 M_{\odot}$ system. These results bring to light the idea that higher BH mass does not always lead to higher disk masses.}}

Regarding the low UCD mass case, the simulations showed very little ability to create accretion disks, \hl{\textbf{especially when comparing identical BH masses for a larger UCD. Both the $M_{UCD}=10^4 M_{\odot}$ with a $10\%$ BH mass fraction (red dashed) and the $M_{UCD}=10^5 M_{\odot}$ with a $1\%$ BH mass fraction (green solid) have identical BH masses ($10^3 \text{ M}_\odot$), only differing in their UCD mass. Even with identical BHs, their capture efficiency is completely different, with the larger UCD system showing, for a large number of AGB stars, $\approx75\%$ of the total gas available is centrally located, and the smaller UCD system only showing $\approx25\%$ of the gas centrally located. This behaviour, along with the observation of lower mass BHs capturing a higher percentage of the gas centrally, implies that the mass of the UCD has a higher effect on the formation of accretion disks than the BH mass. A further discussion on the effects of the BH mass on the formation of accretion disks is conducted in the following Section. With such a large difference between the $M_{UCD}=10^4 M_{\odot}$ and $M_{UCD}=10^5 M_{\odot}$ system's ability to centralise the gas particles, it can be argued that there is a minimum mass required for the UCD such that the proposed mechanism can be effective.}}

\hl{\textbf{On the other side of the mass spectrum,}} the results of the dependence on the number of AGB stars for the high UCD mass regime show a lot of scatter. \hl{\textbf{Also, the previously established trend of an increasing number of AGB stars resulting in a higher percentage of gas mass being centrally located is not present.}} The scatter found can be explained by the increase in SF as the UCD mass increases. From Fig. \ref{fig:agbdep:main}, we see that more stars are bound within $1 \text{ pc}$ for an increasing UCD mass. However, a majority of the total new stellar mass is found outside the bound. Implying that stars form from gas condensing in the accretion disk and then ejected \hl{\textbf{from the central location}} shortly after. \hl{\textbf{Thus, the new stars that form will remove gas from the central disk and potentially disrupt it, as seen in the fiducial model}} (Fig. \ref{fig:best_gasdist2}). \hl{\textbf{This alone does not provide insight into how many stars are forming in the disk and does not fully explain the scatter, as the $M_{UCD}=10^6 M_{\odot}$ also show the same behaviour, in terms of new stars, but produce no scatter in the gas results.}} 

\begin{figure}
    \centering
    \includegraphics[width=1\linewidth]{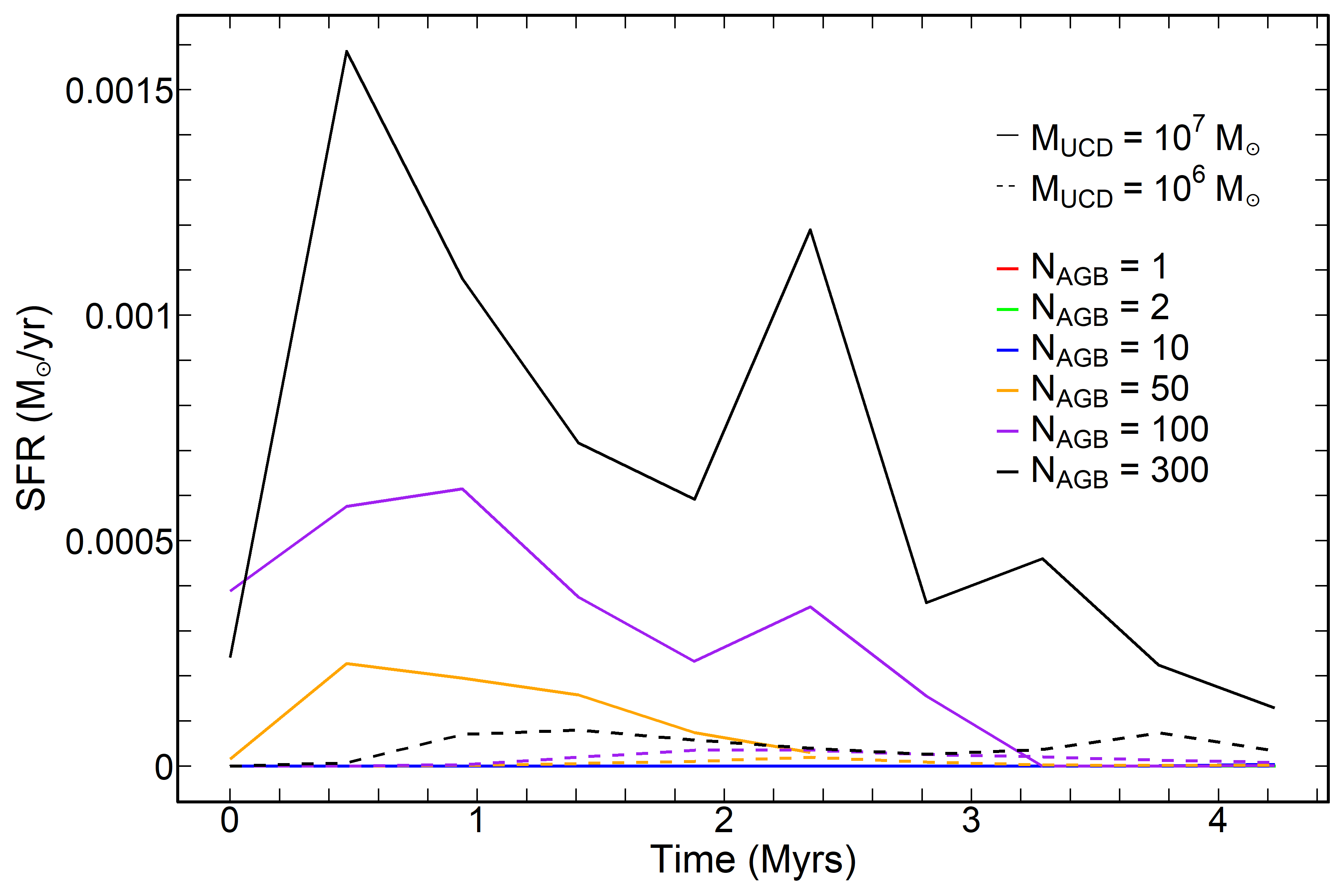}
    \caption{Global SFR ($M_\odot \text{/yr}$) over time for model IDs Ma3 and Ma4. with the following number of AGB stars: $1$ (red), $2$ (green), $10$ (blue), $50$ (orange), $100$ (purple) and $300$ (black). Only the simulations which contained UCDs of mass $10^7 M_\odot$ (solid) and $10^6 M_\odot$ (dashed) are included.}
    \label{fig:agbdep:newstar}
\end{figure}

\hl{\textbf{Fig.}} \ref{fig:agbdep:newstar} \hl{\textbf{presents the SFR over time for the two high UCD cases to compare how their SF differs.}} The global SFR decreases with \hl{\textbf{a decreasing number of AGB stars, which is a result that can be inferred from the right panel of Fig.}} \ref{fig:agbdep:main}. \hl{\textbf{This is expected, as more AGB stars provide more gas for new stars to form.}} A stronger \hl{\textbf{and more useful observation}} is the difference in SFR for different UCD masses, where there is much more SF occurring over all time for the $M_{UCD}=10^7 M_\odot$ simulations, when compared to $M_{UCD}=10^6 M_\odot$. \hl{\textbf{From previous discussion, SF acts as a regulatory/disruptive mechanism for accretion disk formation and growth. Thus, the higher amount of SF present in the high UCD mass case would make it more difficult for the disks to maintain gas mass in a central location, possibly explaining the scatter observed in the left panel of Fig.}} \ref{fig:agbdep:main}. A possible reason for the increased SF is the added potential from the UCD which concentrates more gas in the accretion disk too quickly and thus forms stars rapidly.

The increased \hl{\textbf{amount}} of SF for larger UCDs, along with the ability of larger UCDs to \hl{\textbf{trap gas in a disk more easily}}, places a constraint \hl{\textbf{for which UCD systems could utilise the proposed mechanism for BH growth. If the UCD has too little mass, then the BH is unable to effectively concentrate gas mass, resulting in little BH growth. However, if the UCD has too much mass, too much concentration of the gas occurs which enhances SF. The rapid SF regulates or completely disrupts the disk, slowing or halting potential BH growth. These two observations together lead to the idea that the proposed mechanism works best for higher mass UCDs where SF is slowed or mitigated by other mechanisms.}}

\subsubsection{BH mass fraction} \label{sec:bhmass}
\begin{figure}
\centering
\includegraphics[width=1\linewidth]{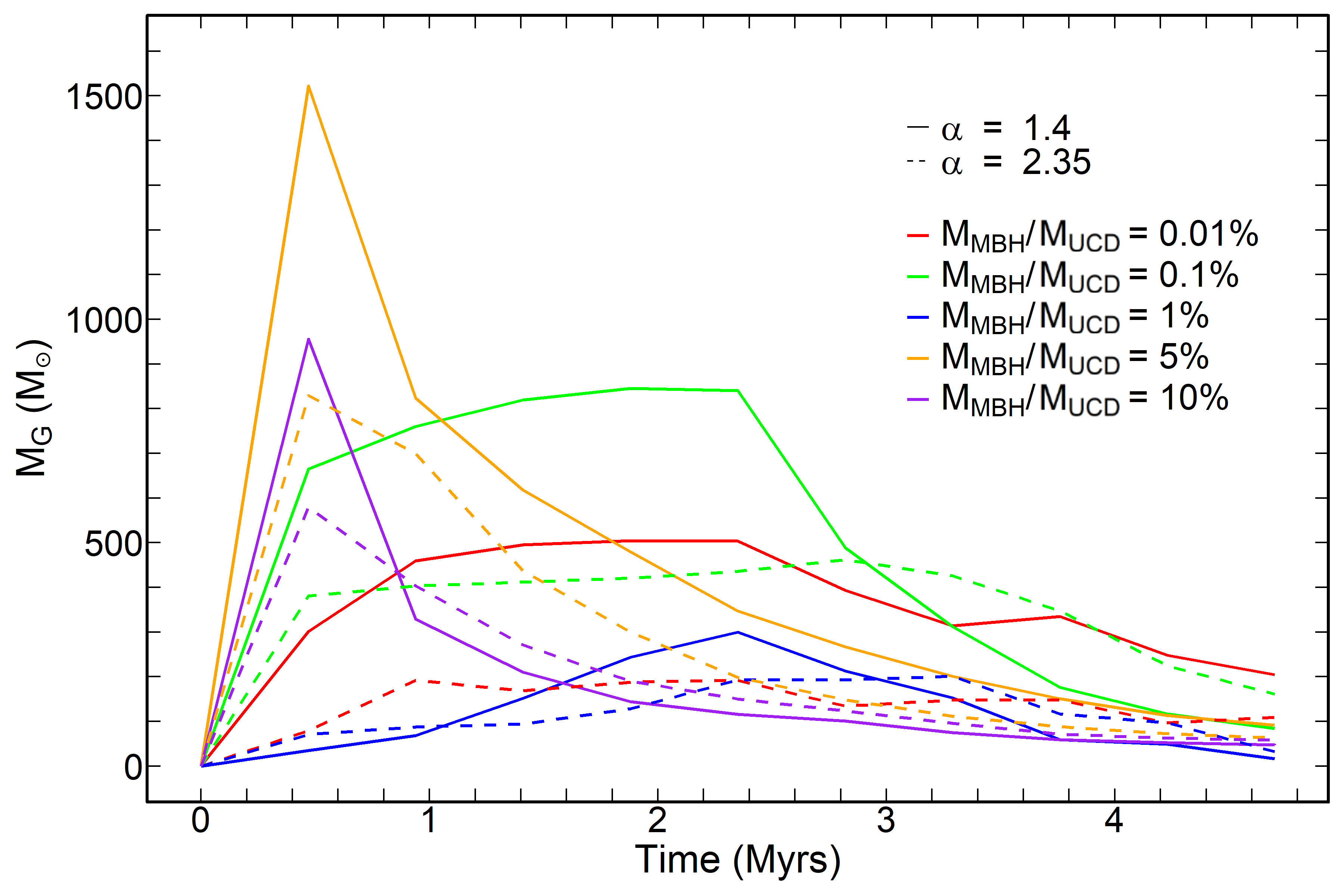}
\caption{Total gas mass ($M_{\odot}$) within $0.1 \text{ pc}$ of the BH over time (Myrs) for several of the realistic simulations for top-heavy (solid) and standard (dashed) along with BH mass fractions of $0.01\% $ (red), $0.1\% $ (green), $1\% $ (blue), $5\% $ (orange) and $10\% $ (purple). These simulations were created using the fiducial model IDs (Mf1 and Mf2).}
\label{fig:realistic:gasvstime}
\end{figure}

The ability of a BH to gravitationally trap gas ejecta is crucial for this model of IMBH growth. In the analysis of how the BH mass fraction affects the model, we utilised model IDs Mf1 and Mf2, which are the fiducial models. Fig. \ref{fig:realistic:gasvstime} plots the total gas mass within $0.1 \text{ pc}$ over the course of the simulations. The gas mass values in the figure were computed by taking the mean of all the simulations that share IMF slopes and BH fractions. The effects of velocity wind and gas temperature were suppressed with this treatment, as their effects get averaged over. As expected, the top-heavy IMF slope shows more mass within $0.1 \text{ pc}$ than the standard IMF slope. With more AGB stars in the simulation, there is more gas mass to trap. However, as we explained in Section \ref{Discussion3}, just the effect of more \hl{\textbf{available gas does not correlate with a higher accretion disk mass. Rather, more available gas mass could result an enhanced SF and thus have the opposite effect.}} 

Fig. \ref{fig:realistic:gasvstime} also highlights that for an increasing BH mass fraction, the more gas mass is trapped. Comparatively, this trend is not as strong, as we observe that the $10\%$ BH mass fraction traps less gas than the $5\%$ for all time. We also see that the smallest BH mass fraction ($0.01\%$) traps more mass within $0.1 \text{ pc}$ for later times. These results \hl{\textbf{directly oppose expectations that}} a greater gravitational potential will trap more gas mass, leading to a larger disk. \hl{\textbf{A hint into why low mass BHs can create more massive disks lies in the profile of the evolution of gas mass near the BH.}} The profile of the \hl{\textbf{results show that the disks around the higher BH mass fractions rapidly gain and immediately}} lose a lot of their \hl{\textbf{gas}} mass early, as is evident by the $5\%$ and $10\%$ BHs. \hl{\textbf{Comparatively,}} the $0.01\%$ and $0.1\%$ BHs show slower \hl{\textbf{disk growth but does not exhibit a massive loss in disk mass}}.

\begin{figure}
\centering
\includegraphics[width=1\linewidth]{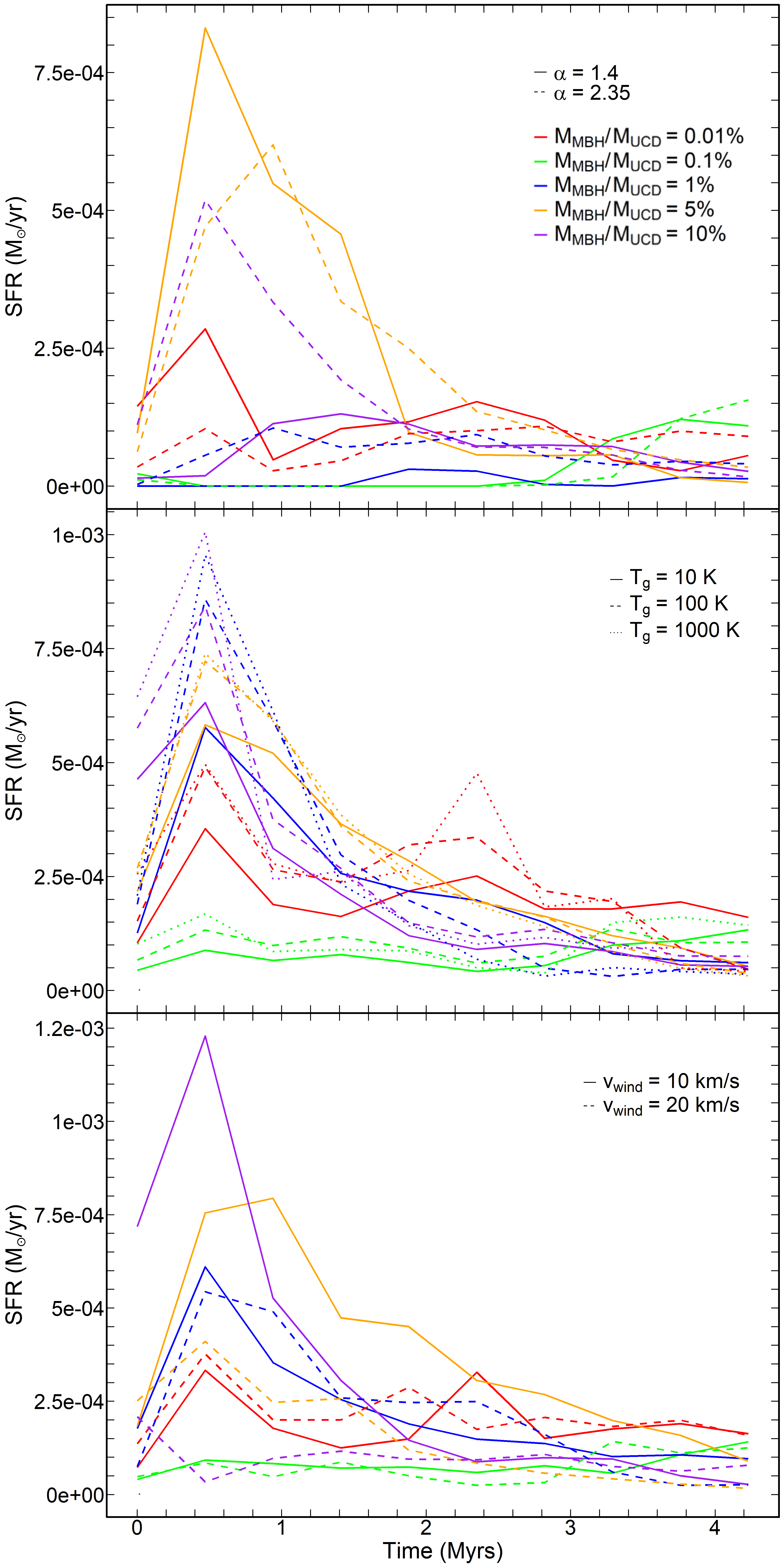}
\caption{Global SFR ($M_\odot \text{/yr}$) of the UCD over time (Myrs) for several of the realistic simulations using models Mf1 and Mf2. Specifically, with BH mass fractions of $0.01\% $ (red), $0.1\% $ (green), $1\% $ (blue), $5\% $ (orange) and $10\% $ (purple). The top plot combines the data of simulations that share an IMF slope of top-heavy (solid) or standard (dashed). The middle plot combines data that share gas temperatures of $10 \text{ K}$ (solid), $100 \text{ K}$ (dashed) and $1000 \text{ K}$ (dotted). The bottom plot considers the velocity of the gas ejecta with values of $10 \text{ km/s}$ (solid) and $20 \text{ km/s}$ (dashed).}
\label{fig:realistic:SFRvstime}
\end{figure}

These differences in behaviour have massive consequences for the evolution of the accretion disk and growth of the IMBH. \hl{\textbf{Similarly to Section}} \ref{Discussion3}, \hl{\textbf{we studied the SFR of these models to determine whether these results could be explained by rapid SF from the larger BH fractions as found previously. Fig.}} \ref{fig:realistic:SFRvstime} \hl{\textbf{plots the SFR over time for the same simulations as Fig.}} \ref{fig:realistic:gasvstime} and uses the same analysis process for the top panel. Regarding the middle and bottom plots, the same data is used, except \hl{\textbf{different simulations get grouped together to cancel out the effect of the other parameters. For example, the middle panel groups simulations based on gas temperature ($T_g$) and the bottom panel groups based on wind velocity ($v_{\rm wind}$).}} When all of the panels are considered, a higher in BH mass fraction results in an earlier and burst-like SF history. For a lower BH mass fraction, the profile of the SFR is more analogous to a constant SFR \hl{\textbf{which is most obvious}} in the $0.01\%$ and $0.1\%$ data. \hl{\textbf{These differences in SF history along with the roll that SF plays in the formation of disks implies that lower mass BHs will not undergo rapid SF which destroys disks. Rather, the amount of new stars produced at any given time is not catastrophic, allowing for the disk to continue. The opposite is true for the large BHs, which exhibit a characteristic burst-like SF, most likely occurring due to the rapid increase in SF, destroying the disk and halting SF. In some cases, the burst of SF is not enough to completely destroy the disk, rather the amount of mass is regulated. This is exactly what is shown in Fig.}} \ref{fig:realistic:gasvstime}, \hl{\textbf{thus the behaviour of SF in these systems well explain the evolution of gas mass in the disk.}}

Looking at the individual panels in Fig. \ref{fig:realistic:SFRvstime}, \hl{\textbf{we can determine how these other chosen parameters affect SF, and in turn the growth of the disk.}} \hl{\textbf{Regarding the IMF slope (top), there does not seem to be a clear trend in regards to how it informs SF. However, in Fig.}} \ref{fig:realistic:gasvstime}, \hl{\textbf{a top heavy IMF slope consistently predicts higher gas near the BH than the standard IMF slope. The lack of influence from the IMF slope implies that the IMF dependence presented in the gas results of Fig.}} \ref{fig:realistic:gasvstime} \hl{\textbf{is purely due to the presence of more gas in the system.}}

\hl{\textbf{Regarding the gas temperature (middle), the higher temperature models produced higher SF rates across all the BH mass fractions for most of the time. This result is unintuitive, as we expect cooler gas to form stars easier even with the cooling mechanisms present in the model. There is a distinction in behaviour for the two regimes of BH mass fraction. For the higher BH mass fractions, the higher gas temperature produces a higher SFR initially, then is overtaken by lower gas temperatures after some time. The lower BH mass fraction regime shows a more constant SFR between the different initial gas temperatures.}}

\hl{\textbf{It is reasonable for the SFR of different initial gas temperatures to show similar values after a long time. This is because the cooling processes allow for the mean temperature differences between models to decrease, leading to similar behaviours after some time. What can not be explained is how the higher gas temperature models show higher SFRs initially. The issues in these results could be resolved by implementing the processes which were not used from Section}} \ref{sec:sim}. \hl{\textbf{For now, this unintuitive result will need to be checked in future work which implement models with more sophisticated mechanisms/processes.}}

\begin{figure}
\centering
\includegraphics[width=1\linewidth]{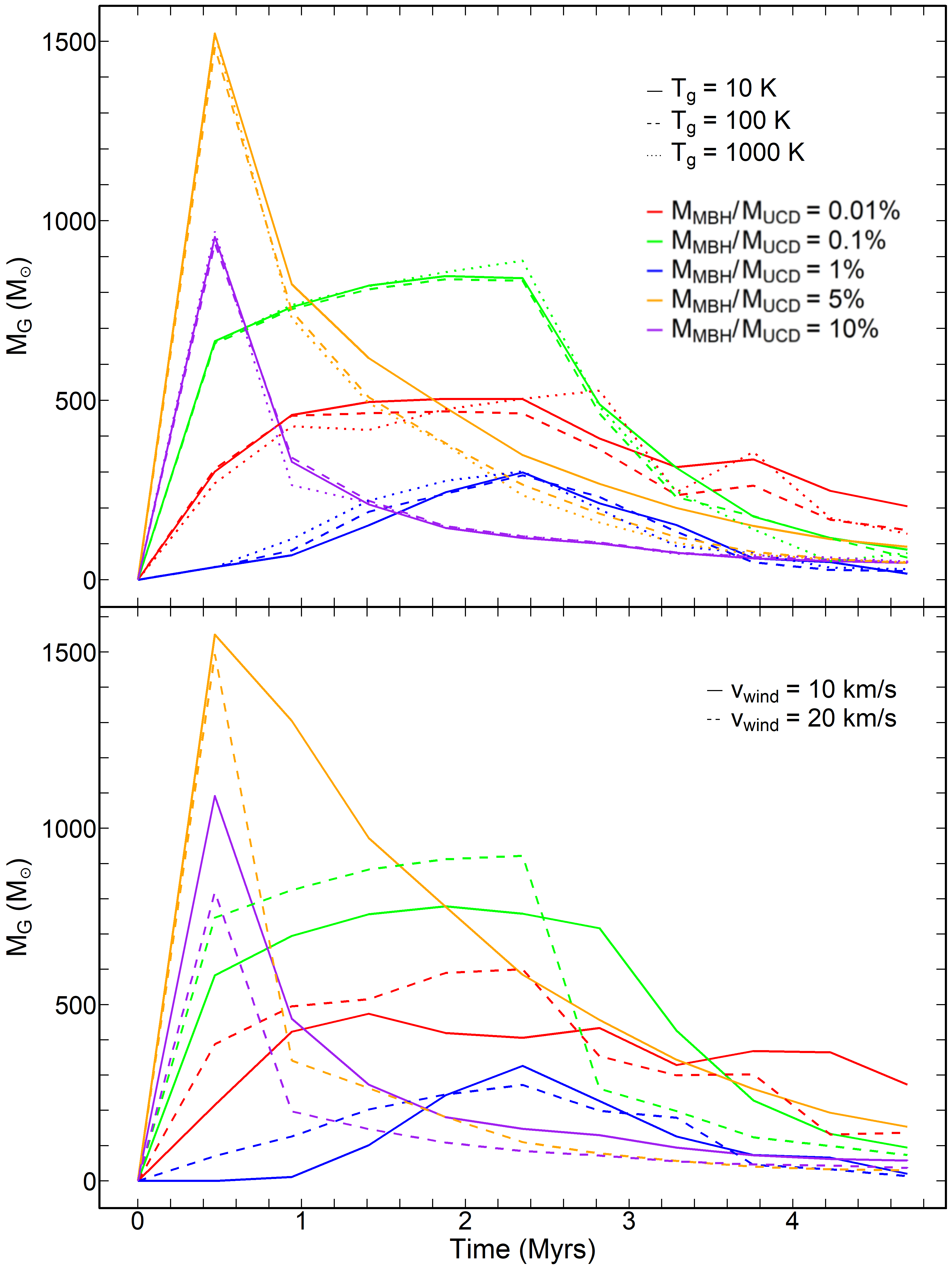}
\caption{Total gas mass ($M_{\odot}$) within $0.1 \text{ pc}$ of the BH over time (Myrs) for several of the realistic simulations using models Mf1 and Mf2. Specifically, for BH mass fractions of $0.01\% $ (red), $0.1\% $ (green), $1\% $ (blue), $5\% $ (orange) and $10\% $ (purple). The top plot considers gas temperatures of $10 \text{ K}$ (solid), $100 \text{ K}$ (dashed) and $1000 \text{ K}$ (dotted). The bottom plot considers the velocity of the gas ejecta with values of $10 \text{ km/s}$ (solid) and $20 \text{ km/s}$ (dashed).}
\label{fig:realistic:gtandvwind}
\end{figure}

With respect to the wind velocity of the gas (bottom), for an increase in wind velocity, the SFR decreases. \hl{\textbf{However, this effect disappears at smaller BH mass fractions, where both wind velocities show similar values for SF.}} Only the \hl{\textbf{BH mass fractions of}} $10\%$ and $5\%$ show discernible differences, \hl{\textbf{with the main difference being the slower wind produces more stars over time}}. The physical reason for these new star results can be investigated in how the gas results are affected by the velocity wind. \hl{\textbf{Fig.}} \ref{fig:realistic:gtandvwind} \hl{\textbf{shows how the chosen parameters of gas temperature (top) and wind velocity (bottom) affect the amount of gas found within $0.1 \text{ pc}$ of the central BH. The data, model ID and analysis process are identical to Fig.}} \ref{fig:realistic:gasvstime}, \hl{\textbf{this time including the mentioned parameters. Considering the top panel,}} the initial temperature of the gas has a negligible effect on the amount of gas trapped. \hl{\textbf{A possible reason for this behaviour is the similar behaviours of SF between the different gas temperatures. As explained previously, the initial gas temperature does not change much about the behaviour of SF in the system, rather it is the BH mass fraction which determines the behaviour. Thus, the initial gas temperature has little effect on the amount of AGB gas which will contribute to BH growth.}}

The wind velocity results have a little more variation. For the lower BH fractions ($0.01\%$, $0.1\%$ \& $1\%$), the velocity that gathered the most gas changes throughout the simulations. Generally, simulations with $10 \text{ km/s}$ gas velocity will trap more gas at the end of the simulation. This is true for the higher BH mass fractions, where the $10 \text{ km/s}$ gas velocity traps more gas at all times.

\hl{\textbf{The wind velocity results show,}} for the lower BH mass fractions, the \hl{\textbf{faster wind case}} can reach the central BH sooner, which explains why there is more mass within $0.1 \text{ pc}$. However, after some time, the increased speed of the gas particles makes it more likely for them to be ejected, \hl{\textbf{as they have a higher kinetic energy}}. This can explain why after some time the simulations with lower velocity winds can retain more gas. From these results, we can see that the temperature of the gas does not affect a central BH's ability to gravitationally trap AGB gas ejecta. However, the velocity of the ejected AGB gas affects not only how fast an accretion disk can form but also how much is lost over time from gas ejection.

\subsubsection{DM halo \& BH position}\label{Discussion2.5}
Previous results and plots were studied without the contribution of a DM halo. Here, we studied model ID Md1, which toggles the DM component and incorporates an additional parameter in the form of the initial BH position from the centre of the UCD ($R_{\rm bh}$). \hl{\textbf{The BH is allowed to move radially after the simulation commences, the $R_{\rm bh}$ parameter only sets the initial position. Expectations of the model are similar to that of the varying UCD and BH mass results, the DM halo's added potential will assist with the trapping of AGB gas in a central location, which would increase potential BH growth. At the same time, SF could be enhanced depending on the speed at which the disk condenses. Regarding how $R_{\rm bh}$ will affect disk formation, the potentials of the UCD/DM halo would not coincide with that of the BH, which could reduce how much gas can be gravitationally bound to the disk. Also, any AGB gas part of the disk could be removed at a later time due to tidal forces.}} 

\begin{figure}
\centering
\includegraphics[width=1\linewidth]{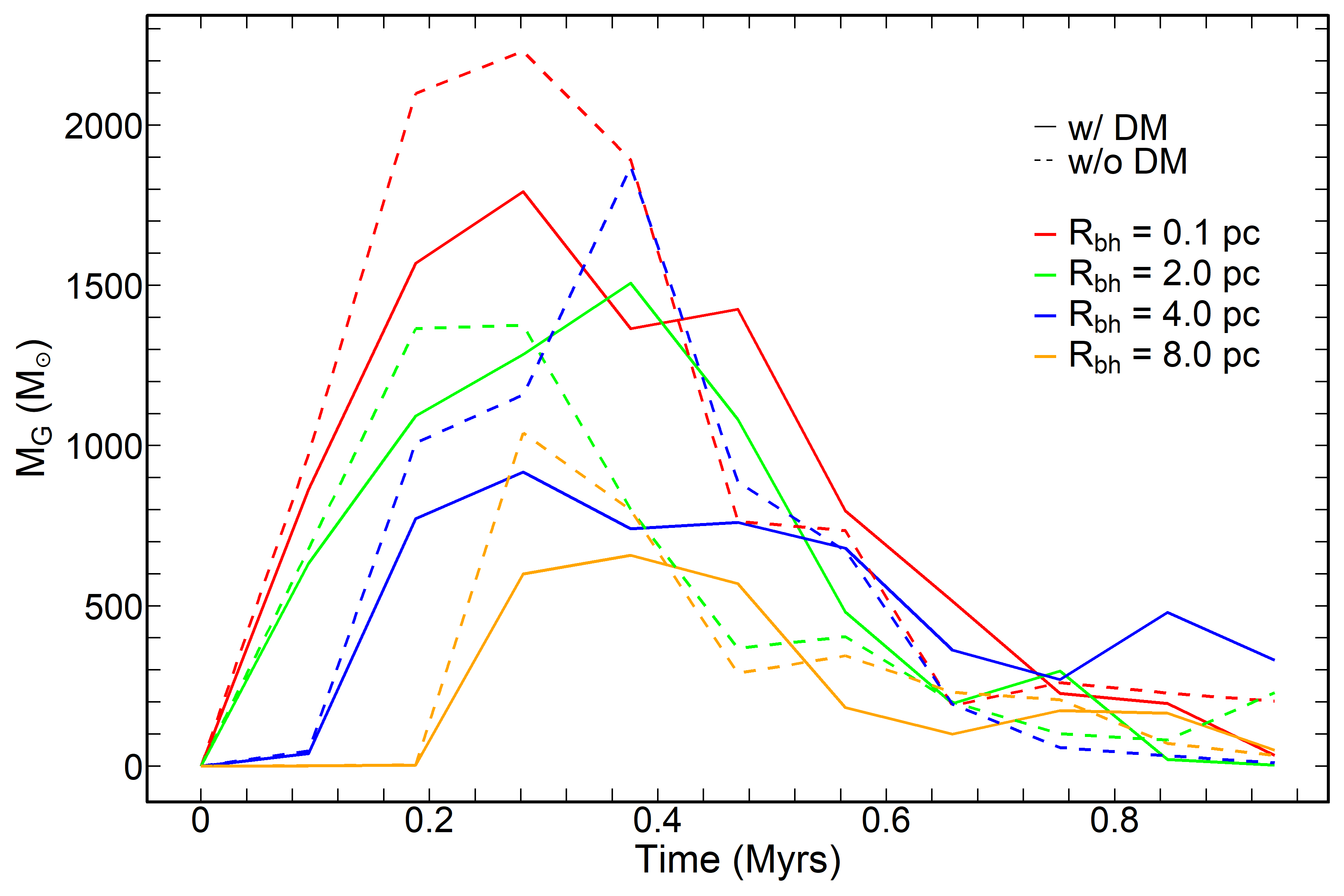}
\caption{Total gas mass ($M_{\odot}$) within $0.1 \text{ pc}$ of the BH over time (Myrs) for different BH positions relative to the centre of the UCD ($R_{\rm bh}$). The model ID for these results are Md1 with the added position parameter included values of $0.1 \text{ pc}$ (red), $2.0 \text{ pc}$ (green), $4.0 \text{ pc}$ (blue), $8.0 \text{ pc}$ (orange). Also shown are the simulations that included DM (solid) and excluded DM (dashed), with the DM simulations placing the UCD initially at $100 \text{ pc}$ from the centre of mass of the DM halo. The simulations used a $5\%$ BH and an IMF slope of $\alpha=1.4$.}
\label{fig:realistic:MBHposition}
\end{figure}

The question of whether UCDs possess DM haloes is still an active area of study. Some evidence explains the elevated M/L ratios of UCDs as a consequence of large DM haloes \citep{baumgardt2008high} and the tidal stripping formation mechanism. However, specific UCD studies using internal stellar kinematics show no evidence of a DM halo \citep{frank2011spatially}. In previous models, $R_{\rm bh}$ has been set to $0 \text{ pc}$, with a typical UCD radius of $10 \text{ pc}$, thus an upper limit is set for the radial distance of the BH \hl{\textbf{since the BH would reside within the system}}. We compare how much gas is present within $0.1 \text{ pc}$ of the BH in Fig. \ref{fig:realistic:MBHposition} which specifically studies the DM contribution and $R_{\rm bh}$. 

The main result shown in Fig. \ref{fig:realistic:MBHposition} is the increase in the gas mass over time with decreasing $R_{\rm bh}$. We also see that the duration in which gas is present within $0.1 \text{ pc}$ of the BH increases with decreasing $R_{\rm bh}$. By comparing the two extremes, the $R_{\rm bh}=0.1 \text{ pc}$ simulation reaches a maximum gas mass of $\approx1750 M_\odot$ and the amount of time the gas mass is present in the disk ranges throughout most of the simulation. Regarding the $R_{\rm bh}=8.0 \text{ pc}$ simulation, the maximum amount of gas at a particular time reached $\approx600M_\odot$ and the amount of time the gas is present decreases to about $0.4 \text{ Myrs}$. This shows a clear relationship between accretion disk growth and the position of the BH within the parent system. That is, for IMBHs to grow via this proposed mechanism, it is more effective if they are centrally located within the UCD. One physical reason for this relationship is that the BH utilises the potential of the parent system when centrally located, allowing for easier gravitational trapping of AGB ejecta. Another less obvious reason for these results is that the less centrally located BHs are closer to fewer AGB stars. This has the effect of limiting the total mass which can feasibly be used for the accretion disk and increases the amount of time it takes for some AGB stars to get close enough to be accreted. This would also explain why BHs further away from the centre start forming accretion disks at later times.

\begin{figure}
\centering
\includegraphics[width=1\linewidth]{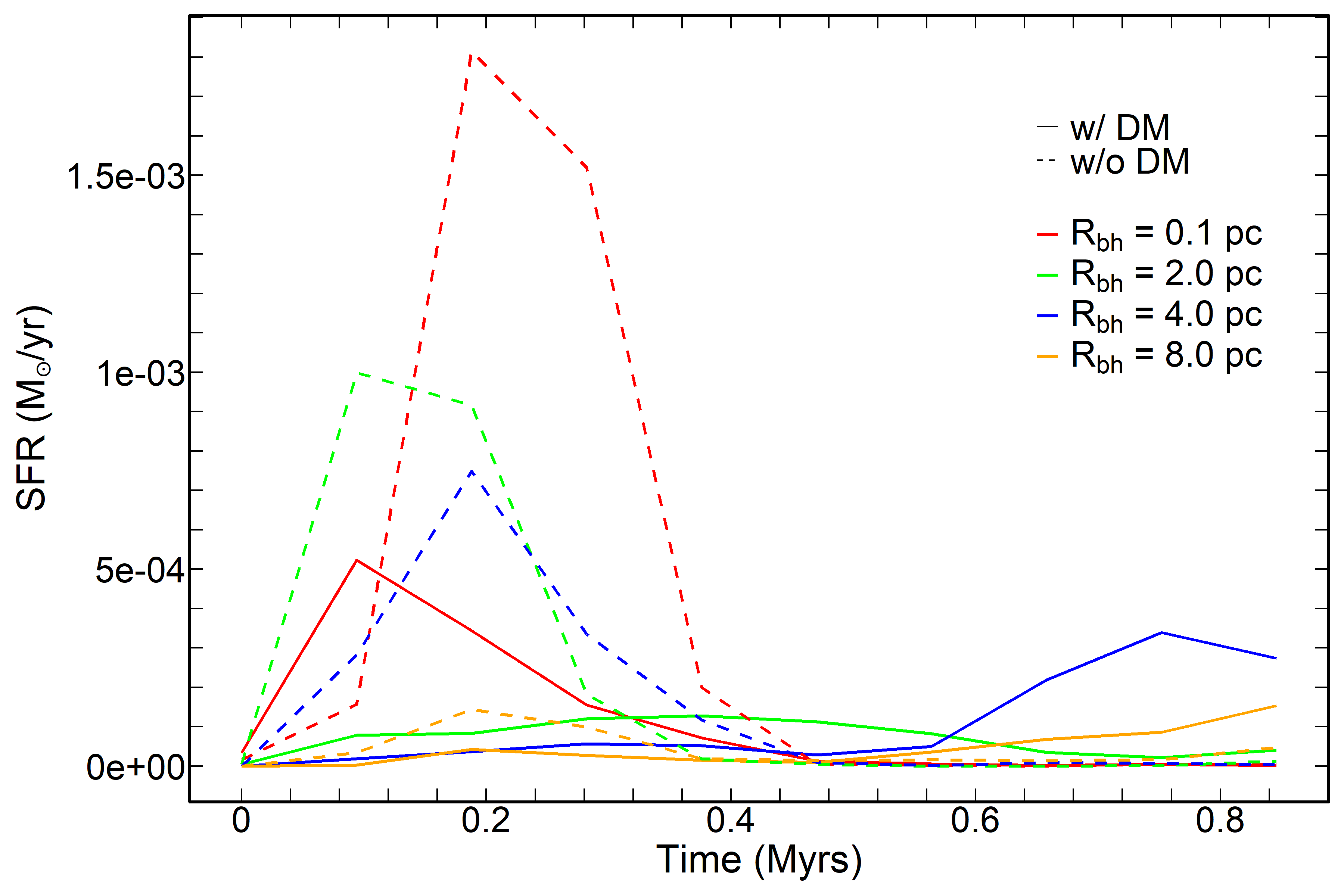}
\caption{Global SFR ($M_{\odot}$/yr) of the UCD over time (Myrs) for different BH positions relative to the centre of the UCD ($R_{\rm bh}$). The model ID for these results are Md1 with the added position parameter included values of $0.1 \text{ pc}$ (red), $2.0 \text{ pc}$ (green), $4.0 \text{ pc}$ (blue), $8.0 \text{ pc}$ (orange). Also shown are the simulations that included DM (solid) and excluded DM (dashed), with the DM simulations placing the UCD initially at $100 \text{ pc}$ from the centre of mass of the DM halo. The simulations used a $5\%$ BH mass fraction and an IMF slope of $\alpha=1.4$.}
\label{fig:realistic:SFRBHposition}
\end{figure}

An unintuitive result appears when considering DM, as we expect that the added potential of DM supports accretion disk growth but the opposite is shown in the results. This behaviour only shows for the early parts of accretion disk formation, as in later times the DM and non-DM simulations generally show similar disk masses. \hl{\textbf{Fig.}} \ref{fig:realistic:SFRBHposition} \hl{\textbf{shows the global SFR for the models presented in Fig.}} \ref{fig:realistic:MBHposition}. \hl{\textbf{The main observation from this plot is that there is more SF occurring for the non-DM models in comparison to the DM models. Therefore, the unintuitive}} results are not due to SF, \hl{\textbf{as we would expect from previous discussions. Another aspect of the SF from these models is that the non-DM results show similar burst-like profiles which were responsible for disk regulation/destruction in previous simulations, as well as the DM model with a central BH (red). The consequences of this observation implies that the presence of DM is slowing down the creation of new stars.}} 

Another possible reason for these unintuitive results stems from the computational necessity to place the UCD $100 \text{pc}$ away from the centre of the DM halo, as the combined potential leads to an overflow of parameters \hl{\textbf{as mentioned in Section}} \ref{sec:sim}. Having the UCD placed off-centre results in the potentials of the UCD and DM halo \hl{\textbf{to be unaligned, possibly resulting in a similar behaviour observed in the varying $R_{\rm bh}$ part of the results. The effect this could have on the AGB gas particles in the DM simulations would be similar as those found previously, specifically, the}} slowing down of gas ejecta, making them reach the BH at later times. This phenomenon is similar to Fig. \ref{fig:realistic:gtandvwind}, which shows that faster gas ejecta will lead to faster accretion disk formation initially, then after some time will show similar accretion disk masses for the slower ejecta velocities. 

\subsubsection{With/Without star formation}\label{Discussion2.55}

\begin{figure}
\centering
\includegraphics[width=1\linewidth]{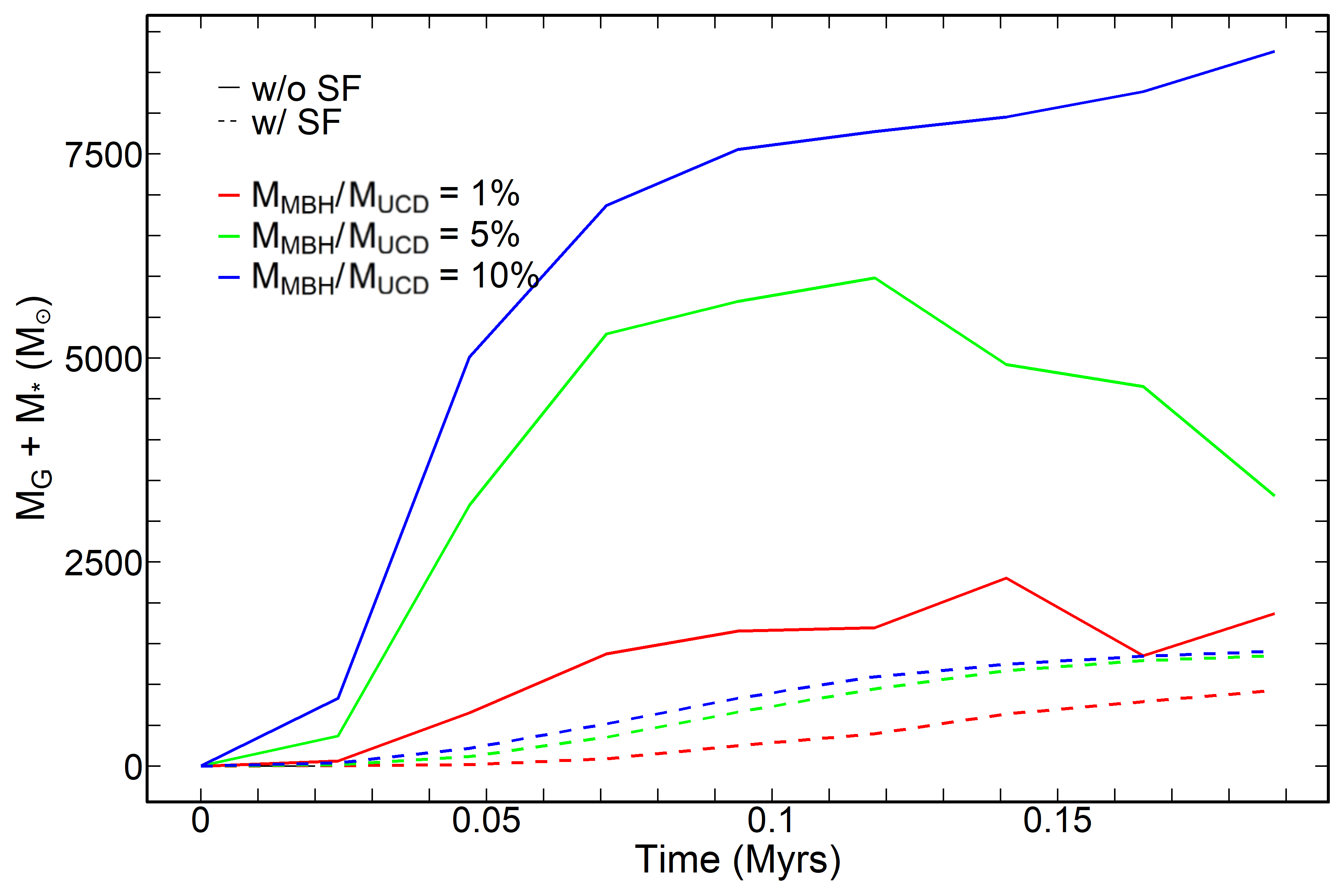}
\caption{Total gas and stellar mass ($M_{\odot}$) within $0.1 \text{ pc}$ of the BH over time (Myrs). The model IDs Ms1 and Ms2 were used to test the proposed mechanism without SF (solid) and with SF (dashed). Other parameters include BH mass fractions of $1 \%$ (red), $5 \%$ (green) and $10 \%$ (blue). The time domain has decreased for higher temporal resolution in the early stages of accretion disk formation.}
\label{fig:realistic:starvsnostar}
\end{figure}

\hl{\textbf{Previous results and discussion has placed SF in an opposing role for the formation and longevity of an accretion disk. It does this by removing AGB gas from the gas by locking it up in stellar mass and by disrupting the dynamics of the disk itself. We have removed the effect of SF for models Ms1 and Ms2 which are compared in Fig.}} \ref{fig:realistic:starvsnostar}, \hl{\textbf{presenting the total amount of AGB gas and new stars present within $0.1 \text{ pc}$ of the central BH. The figure also presents BH mass fraction to explicitly study how the absence of SF can change the previously found behaviour. The reason for this analysis process is so that the only effect of SF is the dynamical disruption, removing the effect of removing gas mass via a change of state. Thus, the}} non-SF simulations (solid) plot only gas mass, while the SF simulations (dashed) plot both gas and stellar mass. 

With this procedure, the non-SF simulations \hl{\textbf{show a larger disk mass}} compared to the SF simulations \hl{\textbf{for all time and BH mass fractions. This effect is much stronger for the higher BH mass fractions, as evident by observing that the difference in disk mass for the two SF cases increases greatly as BH mass increases. This result is expected, as previous results showed that a higher BH mass leads to more SF, resulting in the growth of the disk slowing down. These results further contextualises the role of SF in these systems, especially the high BH mass cases, as just the dynamic effects of new stars forming can regulate the disk mass to a high degree.}} 

\hl{\textbf{For all the BH mass fractions, the rate of growth is also vital, as previous results imply that the speed of disk growth can induce rapid SF. Without the formation of new stars, the speed at which gas mass finds itself within the disk vastly increases, implying that the regulatory effects of star formation begin early into the evolution of the disk. Therefore, SF both reduces the amount of mass in the disk and slows down the growth of the disk. Slowing down the growth of the disk could prevent its complete destruction via rapid SF, as the densities necessary for rapid SF could be avoided.}}

\textbftwo{The nature of a non-SF model is inherently non-physical. However, the use of such a comparison is to test the implementation of the model. This is important for the fiducial model because we wish to check the assumptions made about the gas physics and SF are justified. Appendix \ref{app:newmodel} implements physics that are not present in the fiducial model to test whether they change the results of the fiducial model and to justify the use of it's implementation.}

\section{Discussion} \label{finaldiscussion}
\subsection{Crucial parameters for BH growth}
The proposed method of accretion disk formation has been shown, through these simulations, to provide many examples of accretion disks throughout the parameter space. Such that, we determined which conditions and environments assist or inhibit accretion disk formation. One of the main factors we searched for was the amount of gas ejecta mass within $0.1 \text{ pc}$ of the BH. The fiducial simulations indicated that the disk mass depends highly on the IMF of the parent UCD, \hl{\textbf{as this determines the amount of AGB stars in the system and thus the amount of available AGB ejecta mass.}} While disks can exist in the standard IMF as shown in Section \ref{Discussion2}, a top-heavy IMF \hl{\textbf{produced higher mass disks}} over time. The IMF of UCDs is an active area of study, with some papers attributing the elevated M/L of UCDs to a top-heavy IMF \citep{dabringhausen2009top, dabringhausen2012low, marks2012evidence} and others find evidence for UCDs having a diverse IMF depending on their metallicity or formation history \citep{villaume2017initial, cheng2023initial}. Also, studies have been conducted using recent JWST observations of massive galaxies to constrain the IMF slope of UCDs to the top-heavy regime \citep{bekki2023model}. \hl{\textbf{The potential utility of the proposed mechanism of accretion disk formation and BH growth would depend on the individual IMFs of UCDs.}} 

\hl{\textbf{The initial BH is vital for accretion disk formation, as its properties determine how much growth is required to reach the MBH range and how effective the proposed mechanism is.}} The smallest BHs still created accretion disks of masses comparable to that of the larger BH masses. However, \hl{\textbf{identical}} BH masses found in a low mass UCD showed little to no accretion disk formation, implying that UCD mass \hl{\textbf{and properties are}} more important \hl{\textbf{than that from the central}} BH in creating accretion disks. \hl{\textbf{However,}} observing UCDs is very difficult due to their small size and location within clusters, \hl{\textbf{which make it difficult to determine their properties}}. \hl{\textbf{Because of this}}, the smallest UCD found has a mass of $4.87 \times 10^6 M_{\odot}$ without a confirmed MBH \citep{fahrion2019single}, while the smallest UCD with a confirmed MBH has a mass of $\approx2.63\times10^7M_{\odot}$ in which the MBH constitutes $\approx 8\%$ of that mass \citep{taylor2025supermassive}. The theoretical mass range of UCDs is $10^6-10^8 M_{\odot}$, where the lower bound represents the boundary of the GC regime, and the upper bound represents the boundary between UCDs and dwarf elliptical galaxies. Within this range, and into the GC regime, accretion disks can form with sufficient AGB star ejection. Other simulations have predicted the formation of IMBHs in GCs up to $10^3 M_{\odot}$ \citep{doi:10.1126/science.adi4211}. Our simulations predict that these IMBHs can undergo periods of growth if the system is massive enough, however, if the system is too massive, then we observe that SF is accelerated and the disk becomes disturbed. However, as we explain later, the accelerated SF in reality will occur over larger timescales than simulated, which can spread out SF over a time, making SF cause less destruction in the disk.

We found that SF as a mechanism is the major inhibitor of accretion disk formation and longevity. The formation of stars within the accretion disk both took away gas and gravitationally disrupted the disk. This result explained the scatter found on the left panel in Fig. \ref{fig:agbdep:main} and showed that increasing UCD mass \hl{\textbf{also}} increases SF. When SF was removed, the accretion disks were larger and lasted longer, as evident by Fig. \ref{fig:realistic:starvsnostar}. This showed strong evidence that SF weakens accretion disks or even prevents their formation. Many different factors increase the SFR of the accretion disk, specifically, the mass of the BH affects SF both by total number of stars and how early stars can form. We have shown that for higher mass BHs, the SF \hl{\textbf{showed}} a burst-like structure, resulting in \hl{\textbf{a rapid decrease in AGB gas in the disk. SF can also potentially prevent rapid destruction of the disk as the formation of new stars can immediately slow down the growth of the disk. This slowing down of the disk can prevent the disk from reaching such a large density in a small amount of time, which is the characteristic behaviour of a disk that will rapidly deteriorate from rapid SF.}}

Other positive indicators for IMBH growth through the proposed mechanism can be quantified by high amounts of gas mass in a short distance from the BH for a large amount of time. Both the disk mass and duration of accretion disk activity were found to be highly dependent on the BH position relative to the UCDs centre of mass. This result predicts that IMBHs can form MBHs in these systems more often/easily for centrally located BHs, which was mainly highlighted by Fig. \ref{fig:realistic:MBHposition}.

We emphasised that the interaction between AGB ejecta from separate stars is vital for \hl{\textbf{the proposed mechanism}}. The results of models Ma1-4, and indirectly by fiducial models Mf1-2, highlighted how interactions make it easier for ejecta to fall into the accretion disk of a BH. \hl{\textbf{The effect of which allows for a majority of the gas ejecta particles to become gravitationally bound to the BH. Another mechanism that is important for these disks to become part of direct BH growth is the migration of gas from the outer parts of the disk to the inner parts. Over time, the modelled disks exhibited behaviour which increased the density near the centre and decreased the effective radius of the disk. This mechanism of gas migration allowed for gas particles, which did not lose much energy from the previously mentioned hydrodynamical simulations, to still contribute to future BH growth.}}

\hl{\textbf{The models presented in this study only model a single period of BH growth. We predict that over the several Gyr age of a typical UCD, many periods of BH growth can occur, all contributing to a final MBH that we observe in the universe. For a UCD which starts with a small IMBH of $\approx10^2 \text{ M}_\odot$, it could utilise the proposed mechanism as expected for a small BH mass fraction, growing into a larger BH. Afterwards, if another period of growth occurs, it would now experience the proposed mechanism but analogous to the higher BH mass fraction case. Thus a single system could potentially go through the many different cases of BH growth that was explored in the present study. Such a system could leave evidence of its multiple episodes of BH growth in the form of new stars which formed at their particular period of growth. However, as mentioned previously, a verification of multiple periods of BH growth is presently difficult due to the size, distance and compactness of UCDs.}}

\subsection{Initial BH masses in observed UCDs}\label{sec:initialBHs}

The theoretical mass conversion efficiency of accretion disk mass to BH mass depends on the bolometric luminosity of the accretion disk and the rate of in-falling material into the BH. This efficiency varies greatly with the spin of the BH and the thickness of the accretion disk, \hl{\textbf{the first of which is not possible in these simulations}}. The Novikov-Thorne model for thin accretion disks predicts a radiative efficiency that varies between $\epsilon=6\%-42\%$ \citep{novikov1973astrophysics}, corresponding to BHs with no spin and maximal spin, respectively. Observations have shown similar efficiencies, such as \cite{davis2011radiative}, who used a sample of $80$ pulsars to find a range of radiative efficiencies $\epsilon=3\%-40\%$. Our simulations treated the central IMBH as a point source with no spin, however, the Novikov-Thorne model expects the spin of the BH to evolve over time due to the angular momentum of the accretion disk. Using our simulations, a general approximation \hl{\textbf{for the percentage of the available gas that becomes a part of the disk was found}}. \hl{\textbf{Using this, and a suitable value for the radiative efficiency, the amount of available gas which is converted into added BH mass can be calculated}}. This is how Table \ref{tb:initialbh} was created \hl{\textbf{for different UCDs}}. 

The BH mass fractions in Table \ref{tb:initialbh} range from $6.1\% -31\%$, representing a diverse population of UCDs with MBHs. The last column ($M_{\rm bh_i}$) represents \hl{\textbf{the smallest possible initial IMBH mass that could}} form the observed MBH today using the proposed mechanism. Several UCDs in Table \ref{tb:initialbh} have minimum initial BH masses in the IMBH regime or with negative values. \hl{\textbf{The negative values represent the case where a BH of any mass could utilise the proposed mechanism and still produce the present day MBH mass. The other $M_{\rm bh_i}$ values in the IMBH range are systems which could have had initial IMBHs, however there is a minimum mass constraint for what that mass could be.}} The above consideration used the maximum radiative efficiency ($\epsilon=42\%$) for BHs to limit \hl{\textbf{how much disk mass is converted into}} BH growth. If this parameter was given the freedom to relax to intermediate values, based on the BH spin, then even more UCDs in Table \ref{tb:initialbh} can predict the initial BHs in the IMBH regime. This consideration was used to show that the proposed mechanism can predict BHs initially in the IMBH range from current MBH observations, \hl{\textbf{Whether these systems did in fact use the proposed mechanism requires more observational studies on their stellar populations.}}

\subsection{Origin of high velocity stars in $\omega$ Centauri} \label{sec:omegacent}

\begin{figure}
\centering
\includegraphics[width=1\linewidth]{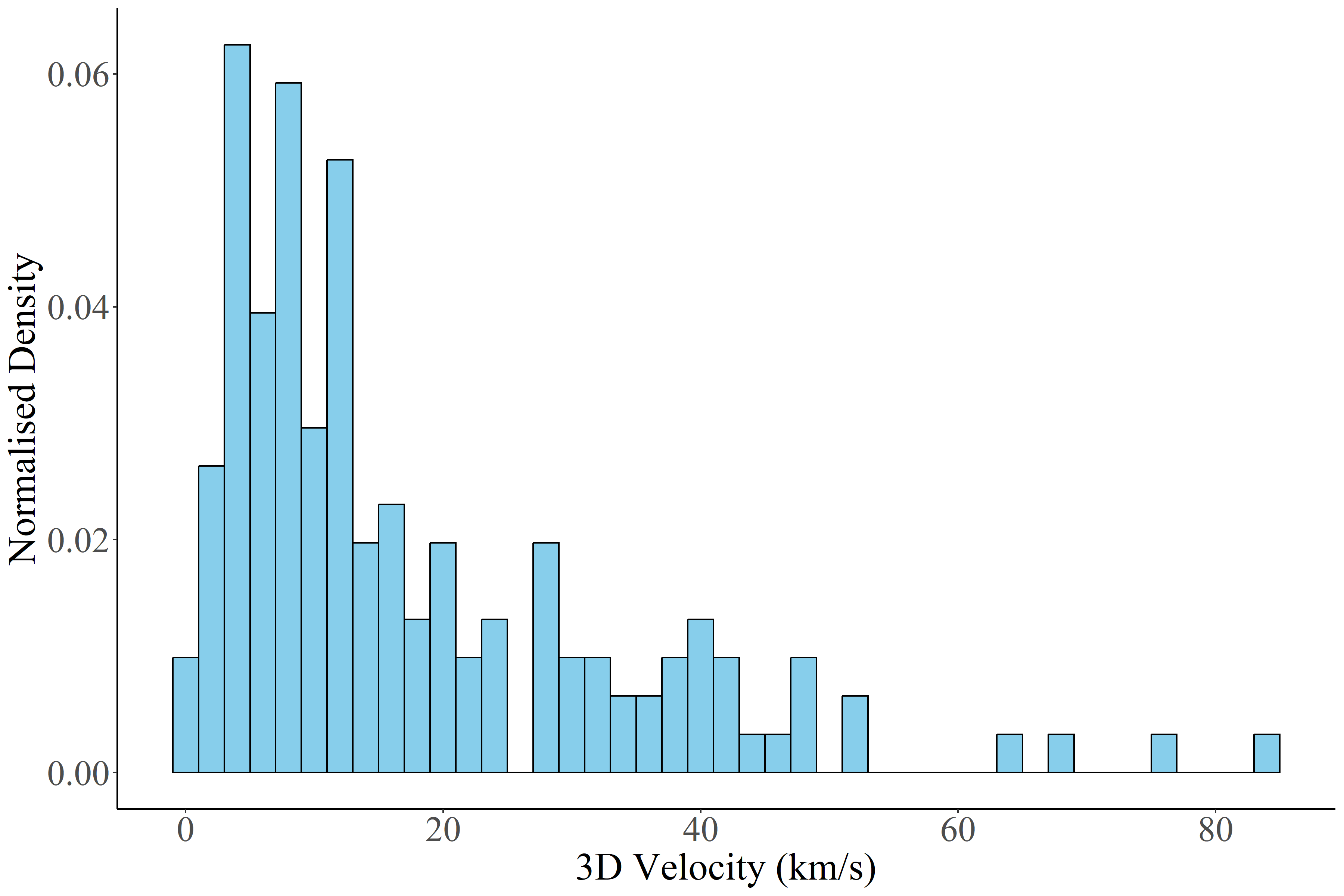}
\caption{Density histogram of the 3D velocity of new stars created in the accretion disk from one of our simulations. The main characteristic shape is a larger population of lower velocity stars with a trail of higher velocity stars. The model ID used for this particular histogram was Mf1 with a BH mass fraction of $5\%$, gas velocity of $20 \text{ km/s}$ and gas temperature of $100 \text{ K}$. High velocity stars are defined as those that have velocities higher than the escape velocity of the system without the presence of a central IMBH. For this particular simulation, the escape velocity at $3 \text{ pc}\approx53.49 \text{ km/s}$.}
\label{fig:discuss:nature}
\end{figure}

The formation of new stars in an accretion disk in these systems could be the origin of high velocity stars in similar systems. Recently, IMBH candidates have been discovered in \hl{\textbf{UCDs and UCD adjacent}} systems. One such candidate was found in $\omega$ Centauri, where stellar dynamics of high velocity stars constrain the mass of a \hl{\textbf{potential}} central IMBH to $\approx 8200 M_{\odot}$ \citep{haberle2024fast}. The few high velocity stars observed would have been ejected without the gravitational influence of such a BH. As seen in the right panel of Fig. \ref{fig:agbdep:main}, most newly formed stars are ejected from the accretion disk and produce orbits that stay bound to the UCD. The velocity distribution of these new stars from one of the simulations produced the histogram in Fig. \ref{fig:discuss:nature}, where we see some high velocity stars in the snapshot. The model IDs used for this purpose were high resolution versions of Mf1 and Mf2. The new stars used for this plot had to satisfy the condition \hl{\textbf{that its total velocity is lower than the escape velocity for every time step. The key observation is that a large population of new stars occupy a low velocity regime, while a few occupy a high velocity regime ($>60 \text{ km/s}$. This kind of velocity distribution was similarly observed in $\omega$ Centauri, in which the high velocity stars provide evidence for a central IMBH.}}

\begin{figure}
\centering
\includegraphics[width=1\linewidth]{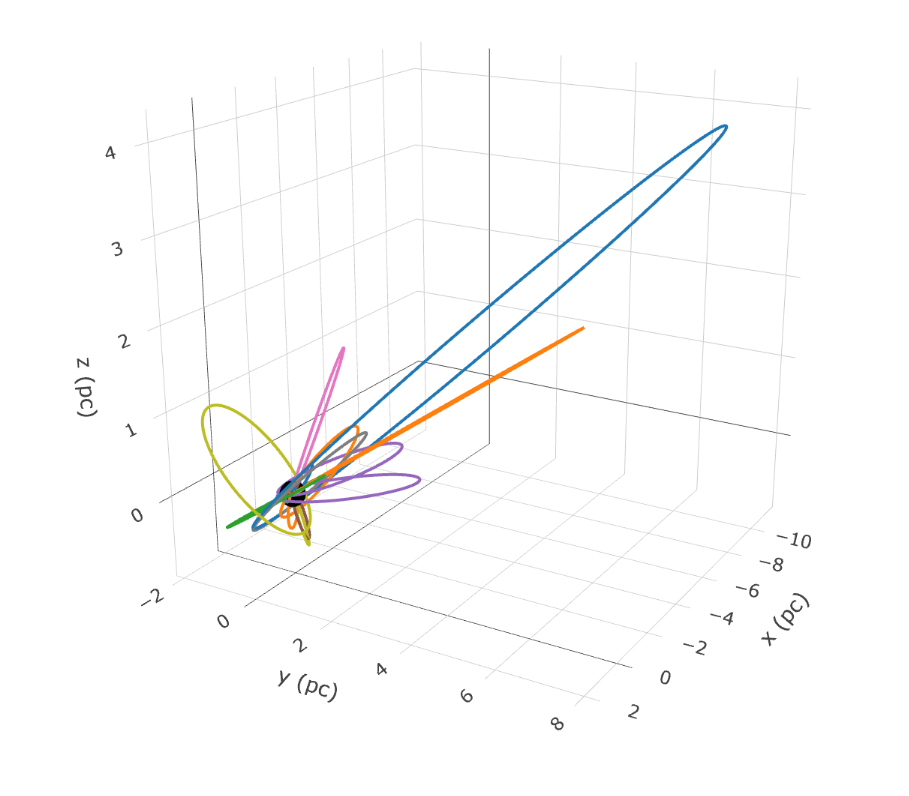}
\caption{3D plot of orbits of new stars that satisfied the condition of being bound to the IMBH. The orbits are calculated via the final time step position and velocities to extrapolate their orbital elements and there are a total of 29 orbits present. The simulation parameters used for this example include; a top-heavy IMF slope, a BH mass fraction of $1\%$ ($10^4 \text{ M}_\odot$), gas temperature of $1000 \text{ K}$ and a velocity wind of $10 \text{ km/s}$.}
\label{fig:discuss:natureorbits}
\end{figure}

The idea behind this comparison was to see whether a random snapshot of the \hl{\textbf{modelled UCD could provide a velocity distribution of new stars which matched that found by}} \cite{haberle2024fast}. \hl{\textbf{Similarities between the modelled new stars and the high velocity stars in $\omega$ Centauri could imply their formation mechanisms are similar. There are several stars in $\omega$ Centauri which are moving faster than the escape velocity of the system (without the influence of a potential IMBH), this is similar for the modelled new stars.}} For Fig. \ref{fig:discuss:nature}, the escape velocity was determined to be $\approx53.49 \text{ km/s}$ for a chosen distance of $3 \text{ pc}$ from the centre \hl{\textbf{of the UCD (no BH influence)}}. The motivation of choosing $3 \text{ pc}$ arises from Fig. \ref{fig:discuss:natureorbits}, which shows that a majority of the orbits reach a maximum distance of $\approx 3 \text{ pc}$. With this value for escape velocity, we can see 4-6 stars which can be considered high velocity in this simulation.

\hl{\textbf{Fig.}} \ref{fig:discuss:nature} \hl{\textbf{was only one of the models that was checked for similar velocity distributions. Out of all the simulations, only high mass IMBHs were able to consistently produce a non-zero number of high velocity stars. This is why Fig.}} \ref{fig:discuss:nature} \hl{\textbf{shows a model with a $5\%$ BH mass fraction. A higher BH mass fraction produces even more high velocity stars. This behaviour implies that the main mechanism responsible for producing these high velocity stars are gravitational interactions between the new stars and the IMBH.}}

The consequences of high velocity stars forming from an accretion disk of AGB ejecta place constraints on the properties of these stars. Firstly, the age of these stars will be younger than that of the general stellar population, \hl{\textbf{which depends on when the period of growth occurred}}. Secondly, the chemical abundance within the high velocity stars will be different from that of the general stellar population. The high velocity stars will form from enriched AGB ejecta and thus be \hl{\textbf{more metal-rich, compared to the older generation}}. In particular, we expect both from models and observations that AGB stars of mass $> 5 M_\odot$ to eject gas rich in nitrogen and form stars with higher [N/Fe] values \citep{ventura2014formation, d2016single}. \hl{\textbf{Future observational studies which aim to determine the chemical properties of the high velocity stars in $\omega$ Centauri could verify}} the prediction that the origin of these stars \hl{\textbf{are from an accretion disk around an IMBH}}. This kind of study would be more difficult for UCDs, but if conducted can verify the proposed mechanism of IMBH growth through the high velocity stars acting as a tracer.

We can view the orbits of these high velocity stars in Fig. \ref{fig:discuss:natureorbits}, which highlights the typical orbits that are expected. The stars that exhibit high velocities near the BH are those with high eccentricities. \hl{\textbf{To find these orbits, we}} utilised the final 3D position and velocities of the stars that satisfy the above conditions to compute the complete orbit around a central IMBH. Here we can see that many of the orbits are close to the central BH, with some extending far in eccentric orbits. This 3D plot further supports the idea that the new stars formed in an accretion disk can escape and form stable orbits \hl{\textbf{all throughout the system}}. With the existence of high velocity stars implying the existence of an IMBH in $\omega$ Centauri and the formation of new stars requiring an accretion disk around an IMBH \hl{\textbf{in the model, these stars could have formed via the same process}}. Thus, a possible formation path for the high velocity stars in $\omega$ Centauri could be through an accretion disk around the IMBH candidate.

There are some caveats that need to be considered with this comparison. First, the distinction between UCDs and GCs as separate objects makes this comparison less useful. Whether these objects should be classified as different objects is still being studied. Second, the work done by \cite{haberle2024fast} consisted of 2D sky velocities, while our simulations allowed for 3D velocities of high velocity stars to be found. Third, the data shown in Fig. \ref{fig:discuss:nature} was found from a higher BH fraction than is expected of $\omega$ Centauri, and we must consider the amount of time covered in the simulations. With more relaxation time, we could see more stars fill out the low velocity regime and a few more high speed outliers. \hl{\textbf{Finally, the stars in the model are ejected from the disk only via gravitational interactions with the central BH. This is because the stars are modelled as collisionless particles with gravitational softening lengths, and thus interactions between stars are limited}} \textbftwo{and unresolved. To elevate this proposition to a proper consideration, future work must investigate these stellar dynamics more accurately and with more sophisticated models than is present in this study. For now, this proposed formation mechanism needs to be studied with a different model from this one.}

Whether UCDs are observed to harbour these high velocity stars is difficult to verify, as these systems are difficult to observe due to their small size and distance. With these caveats, the simulations showed that new stars formed through the concentration of gas in an accretion disk around an IMBH could exhibit high velocities similar to those seen in $\omega$ Centauri. The existence of high velocity stars and their predicted properties could be used as a tracer for whether a UCD with a central MBH previously contained an IMBH and grew via the proposed mechanism. Future observational studies that find this tracer in UCDs presented in Table \ref{tb:initialbh} or others which predict an initial IMBH can be used to verify the proposed mechanism of IMBH growth and to better understand the history of these UCDs.

\subsection{Extra considerations and future work}
The simulations themselves only captured at most $\approx 5 \text{ Myrs}$ of \hl{\textbf{a single period of growth within}} the UCD. \hl{\textbf{The results for a single period of growth can be extrapolated to describe the behaviour of the UCD over several Gyrs.}} \hl{\textbf{Realistically, multiple periods of growth could occur over the age of a UCD or a period of growth could last longer than modelled. Regarding the multiple periods of growth, the outcome of the first would increase the mass of an IMBH then affect the behaviour of the next period of growth. The different behaviours of the accretion disk for different BH mass fractions was discussed in Section}} \ref{sec:bhmass} \hl{\textbf{Regarding potentially longer periods of BH growth, the mechanism of SF would be less detrimental to the accretion disk. As previously discussed, the mechanism of SF has a regulatory effect on the growth of the disk. SF can destroy the disk when there is a rapid formation of new stars when the disk becomes too dense too quickly. If AGB gas is accreted as a slower rate, this would slow the formation of new stars which can allow for disks to last longer even for higher BH masses.}}

In the present study, we did not explicitly model the growth of an IMBH during the simulations, \hl{\textbf{as the resolution requirements for the vicinity of the BH would make it difficult to model the entire UCD.}} The next step in this work will be to incorporate mass growth into the simulation through gas accretion \hl{\textbf{to bridge the gap between the present results and BH growth}}. Another aspect of the model that can be incorporated is the feedback effects from AGNs, as these effects were only considered in the discussion and not the simulations themselves. Adding both BH growth and feedback effects directly in the simulations will provide a more detailed and useful set of results for observational verification. This work can also be continued in a future project that aims to simulate the creation of UCDs through merger events or other processes. The results of such work would provide a clearer and more complete timeline for UCD creation and subsequent IMBH growth into present day observations of MBHs. Currently, we are working on simulations that produce UCDs through mergers between two dwarf galaxies. The results of which will be discussed in a future paper that can be used to supplement the results presented here. Other work involving a more careful consideration of radiative feedback from IMBHs and cases of multiple IMBHs can further show a complete picture. 

\hl{\textbf{Another consideration regarding the present study and discussion is that the total number of AGB stars, gas and properties of the UCD does not take into account the initial dwarf galaxy in which the UCD once resided in. Fig.}} \ref{fig:illustration} \hl{\textbf{highlights that one of the main assumptions of the model involves an NSC which becomes a UCD after the surrounding dwarf galaxy is removed via stripping or tidal interactions. If a period of BH growth occurs while the UCD is still surrounded by a dwarf galaxy, then there could be more AGB stars and thus available gas mass that can contribute to BH growth. The added potential of the dwarf galaxy could also assist in trapping AGB gas near a central BH. Therefore, we predict that these periods of BH growth would be assisted by a surrounding dwarf galaxy.}}

\section{Conclusion}
The first proposal for MBHs to grow via accretion of stellar winds within NSCs was suggested by \cite{norman1988evolution}. However, the formation of accretion disks around IMBHs from stellar winds in these systems has not been clarified yet. Thus, for the first time, we used smooth particle hydrodynamical simulations to test a mechanism for the transition of IMBH to MBHs in UCDs. We proposed that AGB stars within UCDs can have their gas accreted via tidal forces from an IMBH to form an accretion disk. We adopted the assumption that accretion disks would fuel AGN activity and in turn, facilitate BH growth. We also used the Novikov-Thorne model for thin accretion disks when relating the accretion disk mass to increased BH mass. The following results are the main findings of this paper:
\begin{enumerate}
  \item Accretion disks can form around an IMBH through the proposed mechanism of gas ejection of AGB stars. The mechanism involves the interaction of several AGB ejecta from different stars to allow more gas to infall towards the IMBH. We see through density plots and velocity distributions that masses as low as $100 M_\odot$ BH (on the low end of the IMBH range) can form rotating accretion disk structures.
  \item The IMF slope of the system affected the total mass of an accretion disk, with a top-heavy IMF slope predicting more AGBs sooner it could create more massive accretion disks. Also, the position of BHs in the UCD and the UCD in the DM halo were very important for accretion disk growth. The position dictated whether the potentials of the parent objects would work together to support disk growth through shared potentials or inhibit disk growth through tidal interactions. Interactions between stellar ejecta from different AGB stars was important for the formation of the accretion disk. A large amount of the total available AGB gas ($\approx60\%$) was able to be trapped by a central BH for a sufficiently large number of AGB stars and UCD mass.
  \item SF was found to be a large inhibitor of accretion disk growth. The consequences of SF were the removal of gas from the disk and the perturbation of the gravitational structure of the disk. SF highly depended on the mass of the UCD, IMF slope and BH mass. These factors also affected the type of SF occurring, such as burst-like or constant. Accretion disks were destroyed more often from burst-like SF, which can occur less frequently in real systems when gas ejection of AGBs is spaced out and occurs over longer time periods.
  \item We found several UCDs which have MBHs that could have formed through this proposed mechanism of IMBH growth. For these cases, the worst case scenario for radiative efficiency was taken into account and only those with BH mass fractions $>5\%$ were considered. Even with these restrictions, UCDs in the sample were found to either have an initial BH that reached into the IMBH regime or enough gas is present that any initial mass can produce the present day observed MBH. A relaxation of the parameters in the treatment will result in more predictions of UCDs originally harbouring IMBHs that grow via the proposed mechanism. A sample of such UCD candidates with MBHs that form via our proposed mechanism can be found in Table \ref{tb:initialbh}.
  \item By comparing the result of the new star dynamics with velocity distributions of $\omega$ Centauri, we found a similar distribution of star velocities in our simulations. These results provide more evidence for an IMBH in $\omega$ Centauri and provided a possible formation mechanism for the high velocity stars observed. Whether similarly observed high velocity stars are present in UCDs require more stellar kinematic studies to find out, which can be difficult to conduct due to their distance and compactness. If this study can be done, then we can use the predicted properties of these high velocity stars, specifically their enhanced [N/Fe] values, to verify whether they formed via an accretion disk around an IMBH. This, along with candidates for UCDs that form MBHs via the proposed mechanism in Table \ref{tb:initialbh}, can be used to verify this method of IMBH growth observationally.
  \item Future work involving the explicit modelling of BH growth and introducing more complex mechanisms, such as radiative feedback effects, to study their effects on the formation of accretion disks are the next step for this study. Such models will be of greater use for constraining the above results for observational verification of the proposed mechanism. Other types of simulations involving the initial creation of UCDs through mergers and high resolution modelling of the vicinity of the IMBH will fill the gaps in our present study. Such that, a more complete picture of UCD formation and present day MBH observations can be achieved and a more detailed study of the processes involved with translating accretion disk mass to BH growth can be done. 
  \item We showed that IMBHs can have massive accretion disks for a long period of time and that interactions between the IMBH and other stars included more than just gravitational. The gas in the disk can interact with stars passing near the IMBH such that their orbits are influenced via gas-star interaction. In the present simulations, the IMBH was represented by a collisionless particle which is not realistic. Thus, more detailed interaction processes between the IMBH and an accretion disk would need to be implemented in future simulations to construct a more realistic model for these systems with multiple IMBHs. We also plan to investigate these points in future work.
\end{enumerate}

\section{Acknowledgements}

The simulations presented here were performed on the OzSTAR national facility at Swinburne University of Technology. The OzSTAR program receives funding in part from the Astronomy National Collaborative Research Infrastructure Strategy (NCRIS) allocation provided by the Australian Government and from the Victorian Higher Education State Investment Fund (VHESIF) provided by the Victorian Government.

\textbf{We are thankful for the constructive comments provided by the anonymous referee, the effects of which ensured consistency and clarity throughout the paper.}

\section*{Data Availability}

This work makes use of simulations adapted from \cite{bekki2016formation} and \cite{wirth2020formation} which studied the evolution of GCs in dwarf galaxies. Extra additions to these simulations are discussed in Section \ref{sec:sim}. Observational data from \cite{mieske2013central} were also utilised in this work, specifically data involving UCDs with inferred MBHs and their masses. 



\bibliographystyle{mnras}
\bibliography{example} 




\appendix

\section{Testing the Fiducial Model Implementation} \label{app:newmodel}
\begin{figure*}
    \centering
    \includegraphics[width=\textwidth]{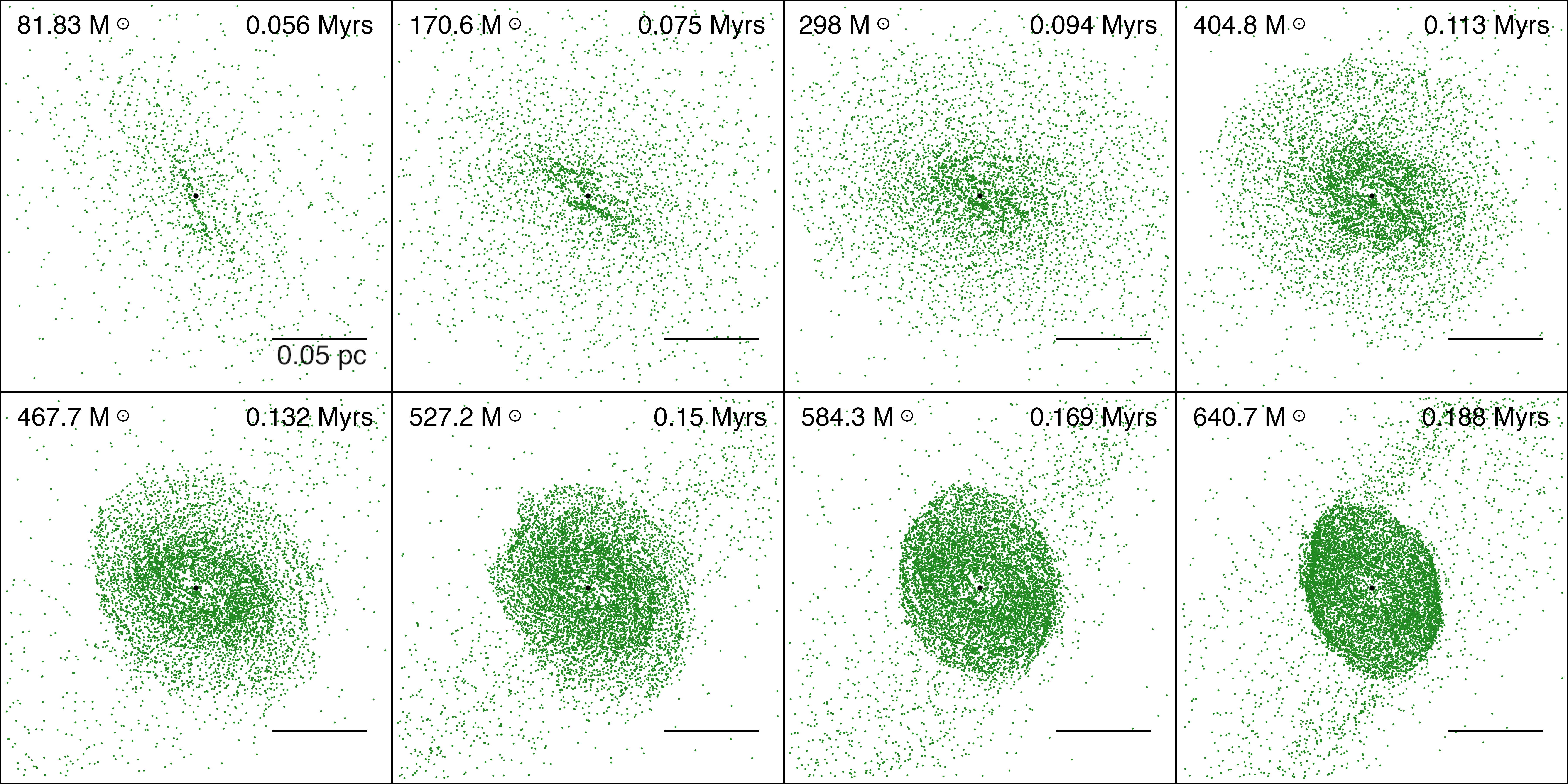} 
    \caption{Snapshots of the gas spatial distribution for the new model. This plot uses the same initial conditions as Fig. \ref{fig:best_gasdist}. The green points represent gas particles from AGB stars and the central black point represents the BH.}
    \label{fig:rev:gasdist}
\end{figure*}

\begin{figure*}
    \centering
    \includegraphics[width=\textwidth]{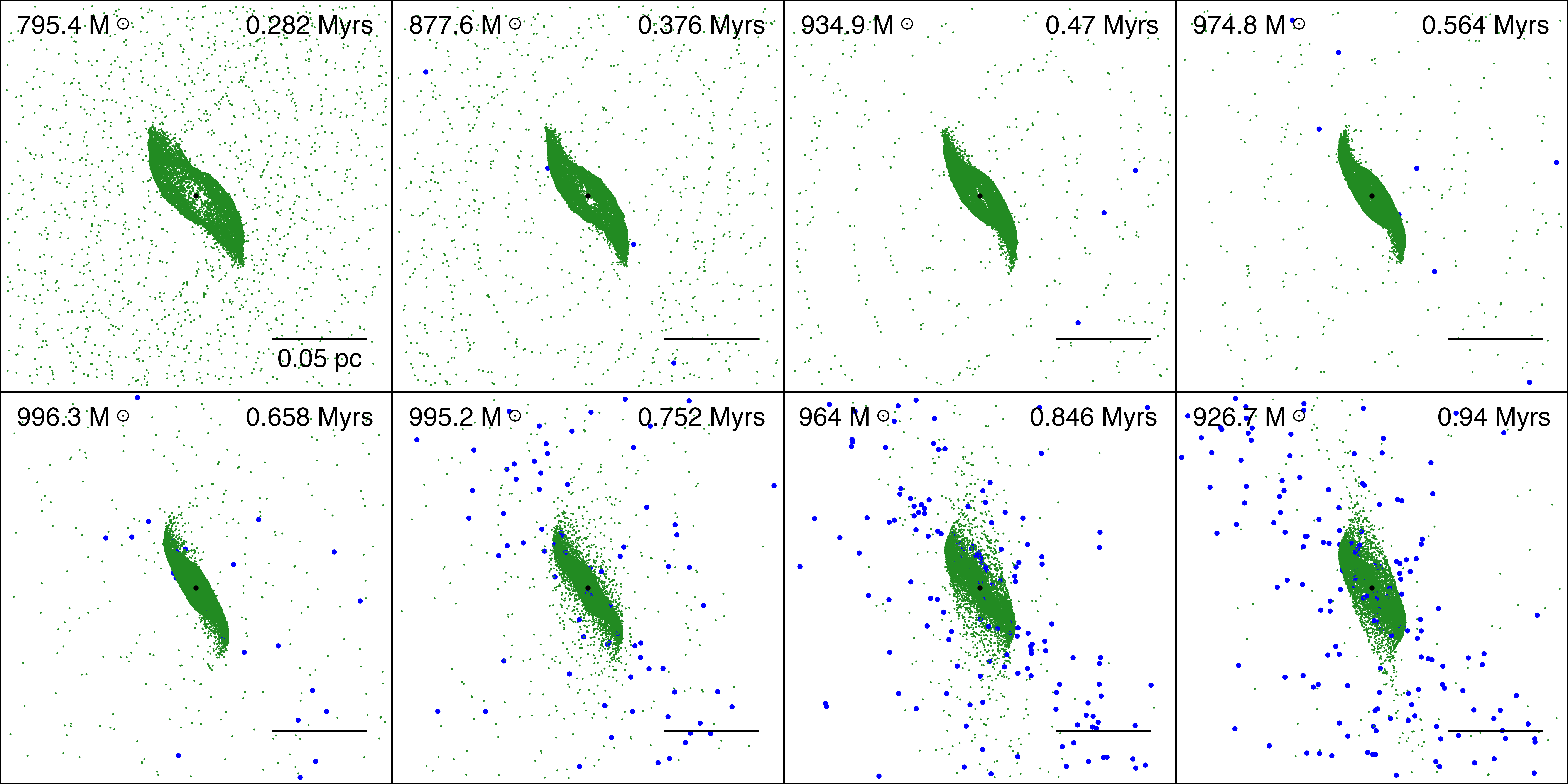} 
    \caption{Snapshots of the gas spatial distribution for the new model. This fig uses the same initial conditions as Fig. \ref{fig:best_gasdist2}. Similarly to Fig. \ref{fig:best_gasdist2}, the green points represent gas particles from AGB stars and the central black point represents the BH.}
    \label{fig:rev:gasdist2}
\end{figure*}

\textbftwo{Here, we present a different model for gas physics to ensure the results are not model dependent and the assumptions made in Section \ref{sec:sim} are sensible. The new model introduces shocks and radiative cooling, along with other details which can be found in \cite{bekki2015dust}. We compare the new results with those from the fiducial model (Section \ref{Discussion2}). Specifically, Figs. \ref{fig:rev:gasdist} \& \ref{fig:rev:gasdist2} are analogous to Figs. \ref{fig:best_gasdist} \& \ref{fig:best_gasdist2}, respectively. The main comparison made between the two models is whether the observed behaviours in the fiducial model is shown in the new model.}

\textbftwo{The new model differs in terms of resolution to include larger gas particles than the fiducial model. The purpose of this is to show that the original packet size of $10^{-2}-10^{-4} \text{ M}_\odot$ have no bearing on the results. Here, the gas packets are $0.28 \text{ M}_\odot$.}

\begin{figure}
    \centering
    \includegraphics[width=\linewidth]{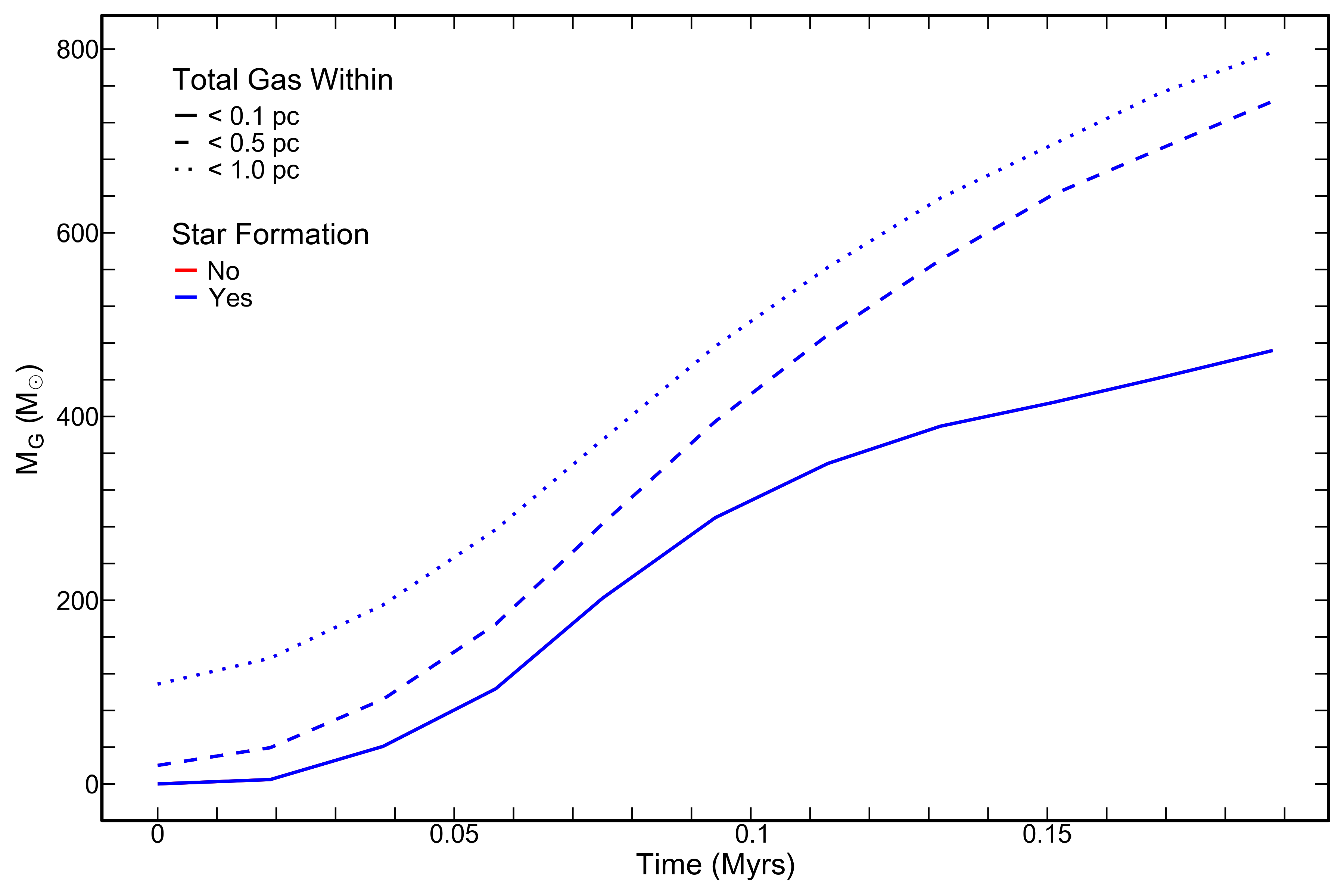} 
    \caption{Total gas pass within radii of $0.1 \text{ pc}$ (solid), $0.5 \text{ pc}$ (dashed) and $1.0 \text{ pc}$ (dotted) from a central IMBH. The parameters used for these results are the same that produced Fig. \ref{fig:rev:gasdist}. Whether the new model allowed for new stars to form is shown in blue and red for with and without SF, respectively.}
    \label{fig:rev:radial_dist}
\end{figure}

\begin{figure}
    \centering
    \includegraphics[width=\linewidth]{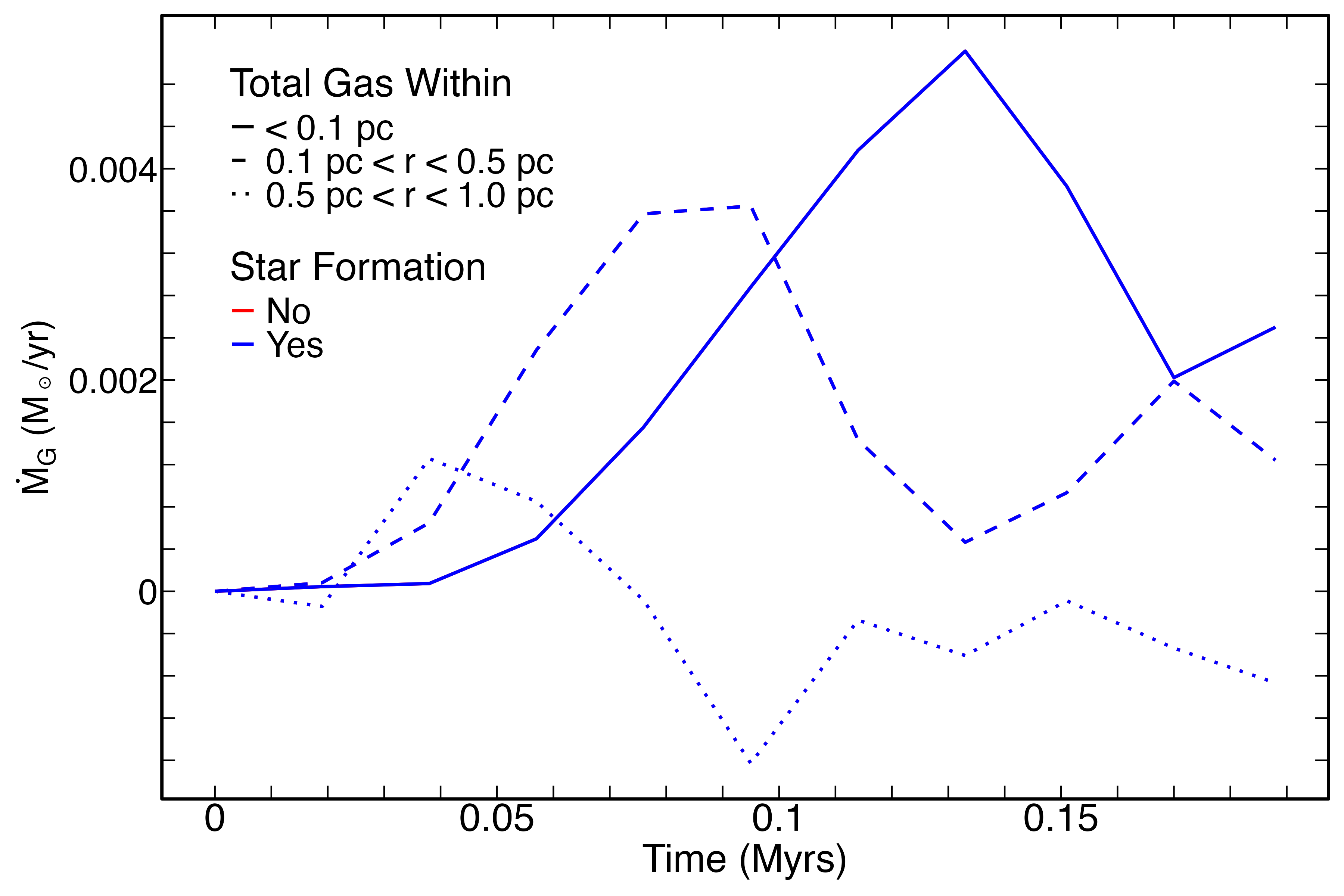} 
    \caption{Accretion rate of AGB gas onto different radial shells. These shells contain the regions $r < 0.1 \text{ pc}$ (solid), $0.1 <r < 0.5 \text{ pc}$ (dashed) and $0.5 < r < 1.0 \text{ pc}$ (dotted). The parameters of the new model are identical to that of the fiducial model and Fig. \ref{fig:rev:radial_dist}, which also shows how allowing (blue) and disallowing SF (red) changes the results.}
    \label{fig:rev:accret_dist}
\end{figure}

\textbftwo{Considering Fig. \ref{fig:rev:gasdist}, the main behaviour is the accumulation of gas mass near the BH. Specifically, the ability for the disk to become more centrally concentrated. This behaviour is better observed in Fig. \ref{fig:rev:radial_dist}, which is analogous to Fig. \ref{fig:best_radialdist}. A difference in presentation of the results are shown, with the process of SF being considered or not as shown by blue and red, respectively. Here, the same trend is observed from the fiducial model. Specifically, the ratio of gas in the outer parts of the disk shrinks with time while the inner parts grow. Thus, the central concentration of gas is observed in both models. Fig. \ref{fig:rev:accret_dist} is analogous to Fig. \ref{fig:best_accrdist} and shows very similar behaviours to the fiducial model. The main similarity is the increase in accretion rate onto the inner shell over time. This, paired with the decrease in accretion rate of the middle and outer shells, further shows the migration of gas that lead to a centrally concentrated disk. These behaviours are also shown in the fiducial model, thus these behaviours do not depend on the model.}

\textbftwo{In Section \ref{Discussion2}, we discussed how SF in the disk causes visible disruptions to the disk. This was process was shown by increasing the BH mass fraction to $5\%$, which was shown to increase SF. Fig. \ref{fig:rev:gasdist2} shows very similar behaviour to Fig. \ref{fig:best_gasdist2}. Specifically, the continued concentration of gas allowing for the rapid formation of new stars. This enhanced SF aligns with a visible disruption of the disk by the end of the model. It is also evident by the slowing and eventual decrease of total gas near the BH as time increases. Thus, similarly to the fiducial model, the new model highlights the role SF plays in the longevity of the disk.}


\textbftwo{The quantitative results from Figs. \ref{fig:rev:radial_dist} and \ref{fig:rev:accret_dist} also show the new model with and without SF to determine whether the choice of modelling SF affects the early formation of a disk via AGB winds. There are no discernible differences between the two models. This result is different to that found in Fig. \ref{fig:realistic:starvsnostar}, which shows that the fiducial model show drastic differences between those that allow and disallow SF. Specifically, the non-SF models can contain much more gas near the IMBH. However, with the new model, both SF and non-SF results reach gas mass values similar to the SF results of Fig. \ref{fig:realistic:starvsnostar}.}

\textbftwo{These results re-contextualises the non-SF results of the fiducial model. The non-SF results were already non-physical, but the new model highlights that the assumptions of an isothermal gas model with no shocks, radiative cooling or SF breaks the model. Even with this new view of the fiducial model, the results of comparing how it handles non-SF conditions help determine where the assumptions of the model become non-physical. Also, future models which include more physics or resolution have a benchmark for how much it improves the physics, as shown with the new model described in this Appendix.}

\textbftwo{The comparison between this new model of gas physics and the fiducial model show that the results above are model independent. Thus, the choice to simplify the model with a isothermal gas physics is can be justified through the little change it makes to the fiducial results. With this initial study into the formation of accretion disks around IMBHs in UCDs, the groundwork has been laid out for future work to build upon, either with more sophisticated models or extended scope.}

\section{Simulation Animation}
We have produced an animation for the fiducial model using 200 snapshots of the gas density distributions, corresponding to $0.94 \text{ Myrs}$ after gas ejection from AGB stars. The animation enables readers to better understand the dynamics of the simulated disks, as well as how the creation of new stars can disrupt but not fully destroy it (shown in teal points). The animation was used with the Mf1 model ID, a $5\%$ BH mass fraction ($5\times10^4 \text{ M}_\odot$), a gas temperature of $10 \text{ K}$ and an initial gas ejection velocity of $10 \text{ km/s}$. Both an edge-on and face-on view is available.


\bsp	
\label{lastpage}
\end{document}